\documentclass[pdflatex,sn-mathphys-num]{sn-jnl}

\usepackage{graphicx}%
\usepackage{multirow}%
\usepackage{amsmath,amssymb,amsfonts}%
\usepackage{amsthm}%
\usepackage{mathrsfs}%
\usepackage[title]{appendix}%
\usepackage{xcolor}%
\usepackage{textcomp}%
\usepackage{manyfoot}%
\usepackage{booktabs}%
\usepackage{algorithm}%
\usepackage{algorithmicx}%
\usepackage{algpseudocode}%
\usepackage{listings}%
\usepackage{aas_macros}
\usepackage{longtable}
\usepackage{array}
\definecolor{customgray}{gray}{0.9} 
\usepackage{subcaption} 
\usepackage[normalem]{ulem}
\useunder{\uline}{\ul}{}
\usepackage{tabularx}
\usepackage{xcolor}
\usepackage[table]{xcolor}
\usepackage{pdflscape}
\usepackage{footmisc}
\usepackage{hyperref}
\usepackage{tablefootnote}
\usepackage{lineno}

\theoremstyle{thmstyleone}%

\theoremstyle{thmstyletwo}%

\theoremstyle{thmstylethree}%

\begin{document}

\title[Article Title]{A Multi-Site Study of Radio Environment for Cosmology Experiments}

\author*[1]{\fnm{Yash} \sur{Agrawal}}\email{yash.agrawal.rri@gmail.com}
\author[1]{Saurabh Singh}
\author[1]{Girish B. S.}
\author[1]{Somashekar R.}
\author[1]{Srivani K. S.}
\author[1]{Raghunathan A.}
\author[1]{Vishakha S. Pandharpure}
\author[1]{Udaya Shankar N.}
\author[1]{Keerthipriya S.}
\author[1]{Mayuri Sathyanarayana Rao}



\affil*[1]{\orgname{Raman Research Institute}, \orgaddress{\street{C. V. Raman Avenue, Sadashivanagar,}, \city{Bangalore}, \postcode{560080}, \state{Karnataka}, \country{India}}}

\date{April 2025}

\abstract{Radio Frequency Interference (RFI) presents a significant challenge for carrying out precision measurements in radio astronomy. In particular, RFI can be a showstopper when looking for faint cosmological signals such as the red-shifted 21-cm line from cosmic dawn (CD) and epoch of reionization (EoR). As wireless communications, satellite transmissions, and other RF technologies proliferate globally, understanding the RFI landscape has become essential for site selection and data integrity. We present findings from RFI surveys conducted at four distinct locations: three locations in India, the Gauribidanur Radio Observatory in Karnataka, Twin Lakes in Ladakh, Kalpong Dam in the Andaman Islands, and the Gruvebadet Atmosphere Laboratory in Ny-Ålesund, Svalbard, Norway. These sites, selected based on their geographical diversity and varying levels of human activity, were studied to assess RFI presence in 30-300~MHz bands, critical for low-frequency observations and experiments targeting the 21-cm CD/EoR signal. Using an automated RFI detection approach via the Hampel filter and singular value decomposition, the surveys identified both persistent and transient interference, which varies with location and time. The results provide a comprehensive view of the RFI environment at each site, informing the feasibility of long-term cosmological observations and aiding in the mitigation of RFI in radio astronomical data. The methods developed to characterize RFI can be easily generalized to any location and experiment.}


\maketitle

\section{Introduction}

Radio Frequency Interference (RFI) has emerged as a major challenge in radio astronomy, particularly with the growing number of radio transmitters both on Earth and in its orbit\cite{inproceedings}\cite{galt1990contamination}\cite{stone1999interference}\cite{2001A&A...378..327F}. The rapid expansion of wireless communication systems, television and FM broadcasts, satellite transmissions, and other Radio Frequency (RF)-based technologies has exacerbated this issue\cite{pankonin1981radio}. International Telecommunication Union (ITU) drafts the international standards, regulations and allocation of these frequency bands\footnote{\label{itufoot}\href{https://www.itu.int/dms_pub/itu-r/opb/reg/R-REG-RR-2020-ZPF-E.zip}{https://www.itu.int}}. ITU defines RFI as the unwanted emissions from devices that violate these regulations. However, in this work, the usage of the term RFI will reflect any anthropogenic signal that contaminates astronomical sky data, regardless of its allocation, as commonly adopted in radio astronomy\cite{2015PASA...32....8O}\cite{2023ApJ...957...78W}.
Choosing a location, therefore, for conducting observations of the radio sky should be preceded by a comprehensive RFI survey of candidate observation sites. Such surveys provide valuable insight into the radio environment, including the nature and extent of the interference. This information is essential for evaluating a site’s suitability for both short and long-term scientific observations.
Such understanding is particularly crucial for experiments that aim to detect faint signals such as the global 21-cm signal  or 21-cm power spectrum from cosmic dawn (CD)\cite{2023JApA...44...10B} and epoch of reionization (EoR)\cite{2013MNRAS.435..584O}. In these experiments, where the target signal is several orders of magnitude weaker than the foregrounds, precise modeling of the data is essential\cite{2017ApJ...847...64M}\cite{2022NatAs...6..984D}\cite{2015ApJ...799...90B}\cite{2019JAI.....850004P}.
For such precise modeling, the quantification of potential data contamination due to RFI becomes an essential parameter\cite{2020MNRAS.498..265W}.
The quality of the data depends on various site characteristics, including the RFI environment, horizon profile\cite{2024MNRAS.527.2413P}\cite{2021ApJ...923...33B}, and accessible foreground sky\cite{761c98bcdd13467f87dd40e7180307b3}. Among these, RFI determines the usable frequency band, while the horizon profile and sky foregrounds play a crucial role in shaping the data modeling. As a case study, we take the Shaped Antenna measurement of the background RAdio Spectrum (SARAS) experiment, which is one such radiometer experiment aiming to detect the elusive global 21-cm signal\cite{t2021saras}\cite{2018ExA....45..269S}\cite{2013ExA....36..319P}. It uses a monopole antenna with a lifted toroidal beam that has a null at the zenith\cite{2021ITAP...69.6209R}. The antenna is deployed in a radio-quiet zone to minimize RFI, and the system is optimized to enable separation of the cosmological signal from foregrounds based on their differing spectral characteristics. The frequency band of interest in which this signal lies is 40-200~MHz\cite{2012RPPh...75h6901P}. Multiple frequency channels in this band are used for terrestrial and satellite communications. Besides that, the horizon profile significantly affects the RFI environment and foreground modeling\cite{2024MNRAS.527.2413P}. Hence, a detailed study of the site from the above perspective is crucial before any deployment.

\setlength{\parskip}{1em}

This paper discusses the radiometer design that has been developed for a robust RFI survey for the SARAS site selection. Along with the system design, we also discuss the associated calibration methods and RFI detection algorithms to assess a site's suitability for precision sky measurements. It presents the results of surveys conducted at four diverse locations: the Gauribidanur Radio Observatory in Karnataka (GRO), the Twin Lakes in Ladakh (TLL), the Kalpong Dam in the Andaman Islands (KDA) (all locations in India), and the Gruvebadet Atmosphere Laboratory in Ny-Ålesund, Svalbard, Norway (GLS). These locations were selected due to their varying levels of human activity, geographic isolation, and proximity to urban centres. The surveys targeted frequency bands 30-300~MHz, which include the range of interest for experiments aiming to detect the global 21-cm signal from the cosmic dawn and epoch of reionization\cite{2012RPPh...75h6901P},\cite{1999A&A...345..380S},\cite{2006PhR...433..181F},\cite{2004MNRAS.347..187F}. 
This paper is organized as follows:
Section \ref{sec: Instrumentation} provides a detailed overview of the instrument used for the RFI survey, including the complete RF chain and antenna. Moving on to Section \ref{sec: Observations}, we describe the surveyed sites, their land topology, and accessible foregrounds. This section also outlines the calibration techniques applied to the observed data, leading to calibrated drift scan plots from the survey and their major features. 
In Section \ref{sec:Data_analysis}, we employ the algorithms to quantify these observations in terms of RFI. In Section \ref{sec: Results and Morphology of RFI}, we present the metrics such as morphologies and occupancies of the RFI detected at these locations. Finally, Section \ref{sec:Discussion and Conclusions} compiles the conclusions drawn from the survey and analysis, offering a broader discussion on their implications. While these results are applicable to any precision experiment operating within the 30–300~MHz frequency band, we discuss the impact of these results on the deployment of SARAS as a case study.

\section{Description of the RFI radiometer setup} \label{sec: Instrumentation} 
\begin{figure}[htb]
    \centering
    \includegraphics[width=0.8\textwidth]{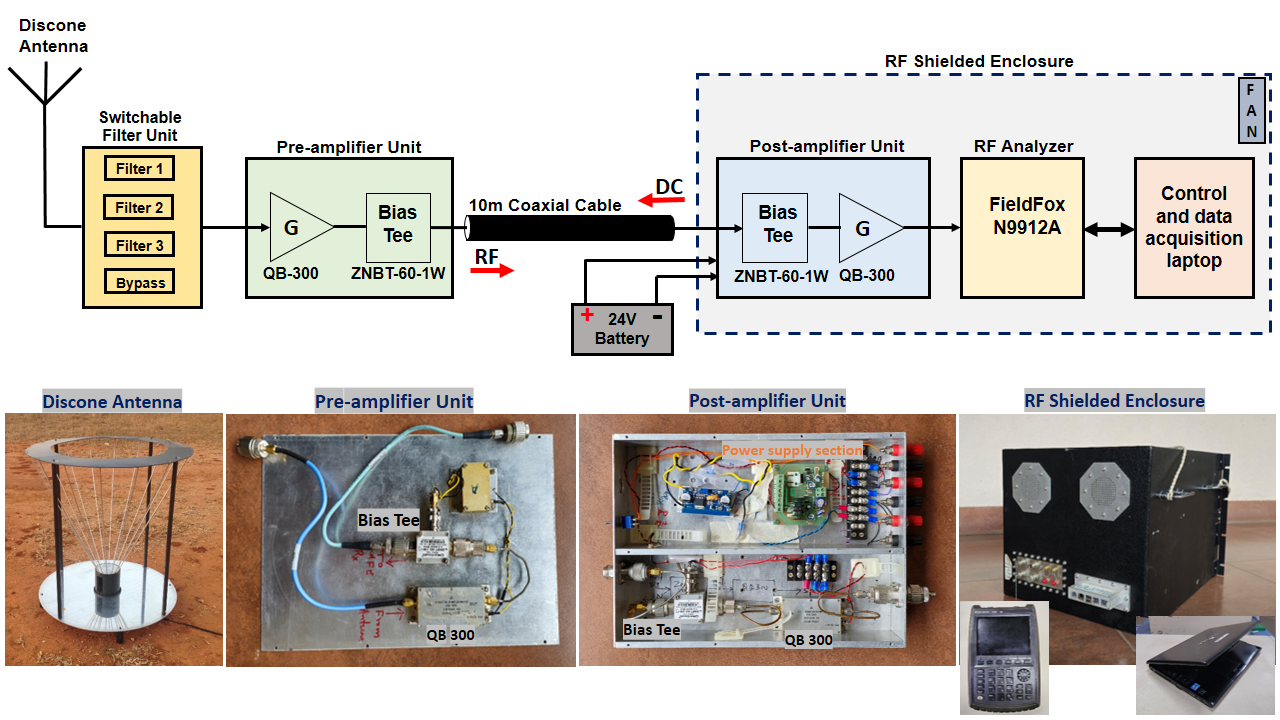} 
    \caption{The top panel shows a block diagram of the RFI radiometer setup. The pre-amplifier module amplifies the weak RF signal from the discone antenna. The post-amplifier unit, RF analyzer, and data acquisition laptop are housed inside an RF-shielded enclosure (shown within a dashed box). Apart from pictures of the discone antenna and the RF-shielded enclosure, the bottom panel shows an open view of the internal circuitry of the pre-amplifier and post-amplifier unit.}
    \label{fig:Block_diagram}
\end{figure}

Radio-frequency characterization of a candidate site is carried out by deploying a portable RFI radiometer setup. Figure \ref{fig:Block_diagram} shows a block diagram of the portable RFI recording setup capable of continuously recording RF spectrum in the 30-300~MHz frequency range. It consists of a discone antenna, which can be easily assembled in the field. It is followed by front-end electronics consisting of a switchable filter unit and a pre-amplifier module to amplify the analog signal at the antenna output. The back-end electronics, connected via 10~m low-loss coaxial cable, are housed inside an RF-shielded enclosure. It consists of a post-amplifier module, an RF analyzer, and a control and data acquisition laptop.  An RF analyzer produces a spectrum of the input RF signal. The laptop controls and acquires data from the RF analyzer via an Ethernet cable. 24~V DC supply from batteries powers the entire RF signal chain.

\subsection{Discone Antenna}
\begin{figure}[H]
    \centering
    \includegraphics[width=0.8\textwidth]{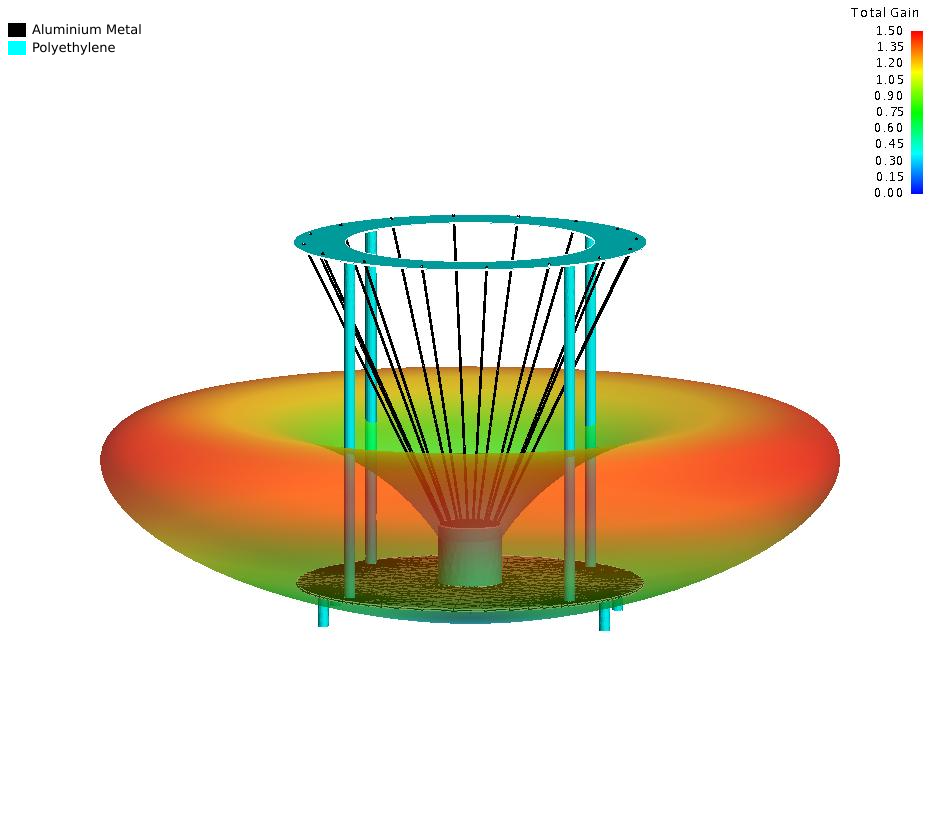} 
    \caption{Geometry of the discone antenna as simulated in FEKO. The beam pattern shown is at 150~MHz.}
    \label{fig:Discone antenna geometry}
\end{figure}

 A discone antenna is used to convert the incident electromagnetic waves into electrical signals. Figure \ref{fig:Discone antenna geometry} shows the geometry of the discone antenna as simulated in FEKO\footnote{\href{https://altairhyperworks.com/product/FEKO}{Altair FEKO}}. FEKO is an Electromagnetic simulation software that includes frequency domain solvers like Method of Moments, Finite Element Method, etc, as well as Finite Difference Time Domain, which is a time domain solver. To simulate the RFI measurement antenna, we have used the Method of Moments solver, which employs a surface meshing technique. We approximate a biconical antenna, replacing one of the two cones with a disc. It is an omnidirectional antenna with a wide bandwidth. 
Figure \ref{fig:Discone antenna geometry} also shows the 3D primary beam of the antenna at 150~MHz. The primary beam closely resembles the SARAS beam\cite{2021ITAP...69.6209R}, with nulls along horizon and zenith. With significant response at low elevations, these antenna systems are prone to receiving high RFI, which is usually transmitted along the horizon. The discone antenna is optimized in the 30-300~MHz frequency range for efficient performance. The cone is realized with multiple spokes or rods to reduce weight and wind resistance. This also allows full assembly of the antenna within a few minutes.

Sixteen rods, each with a diameter of 6~mm and a length of 94~cm, are mounted on a plate with a diameter of 170~mm and a thickness of 6~mm. A circular disc with a thickness of 6 mm and a diameter of 1~m acts as the ground plane for the antenna.
A PVC support structure is also a part of the antenna, consisting of 4 pegs at the bottom, a ring at the top to maintain the metal rods in place, and 4 PVC rods that support the ring.

\begin{figure}[H]
    \centering
    \includegraphics[width=0.8\textwidth]{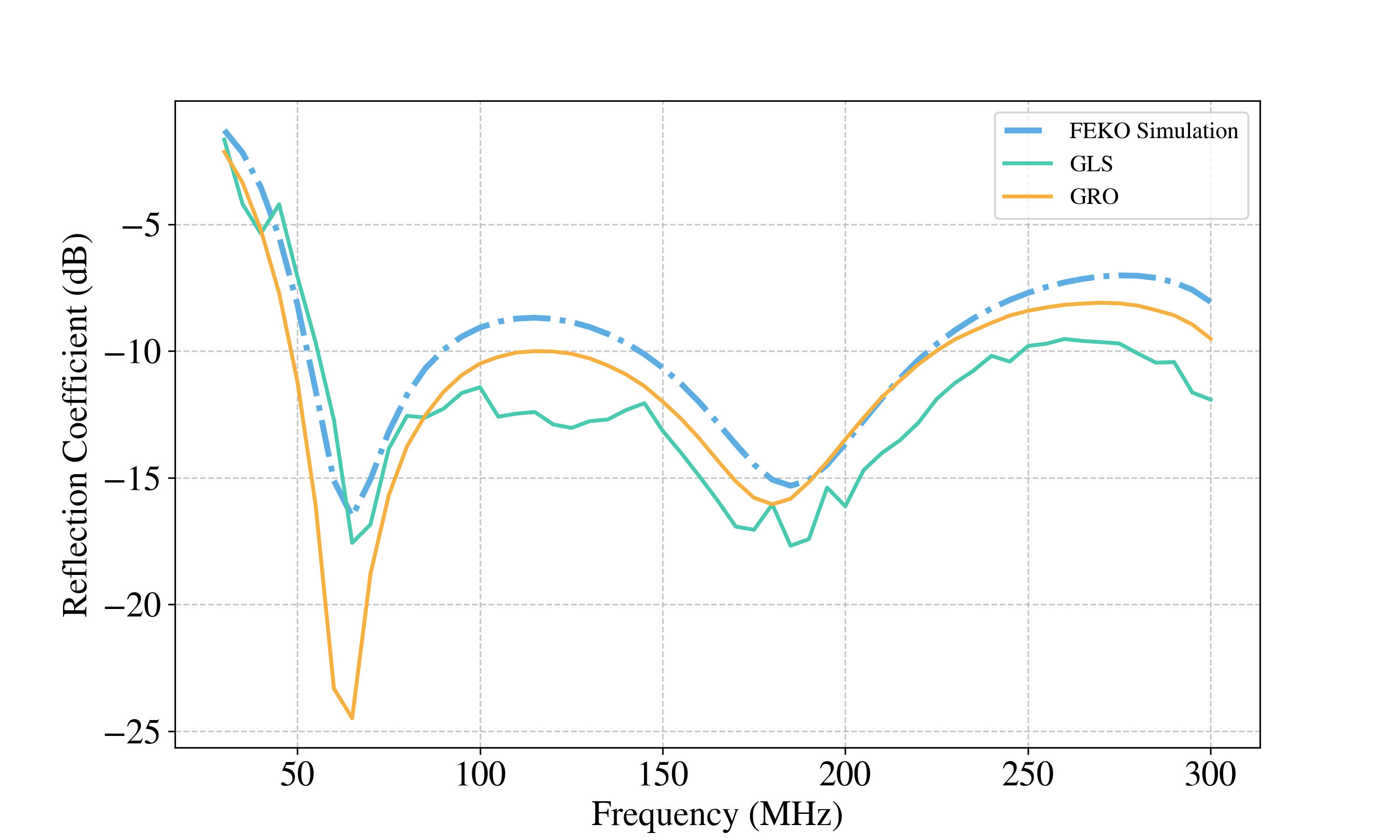} 
    \caption{Comparison of the reflection coefficient of the antenna with simulations in FEKO: blue dashed line shows simulation with realistic ground profile, orange shows the measurement at GRO, and green represents measurement at GLS. The GRO profile is less noisy than GLS because of the integration times of measurements.}
    \label{fig:Measured and simulated S11 of the discone antenna}
\end{figure}

The antenna reflection coefficient, governing coupling of available power at the antenna into the receiver\cite{5162049}, varies from -~3~dB to -~25~dB over the frequency range of 30-300~MHz. Figure \ref{fig:Measured and simulated S11 of the discone antenna} shows the reflection coefficient of the antenna measured at GRO and GLS along with simulated S11 with FEKO. For the simulation, the antenna is modeled 10~cm above an infinite ground plane, with a dielectric constant of 5 and a conductivity of 0.0022~S/m, representing dry ground conditions\cite{2021ITAP...69.6209R}. 

\subsection{RF Receiver Chain}
\begin{figure}[H]
    \centering
    \includegraphics[width=0.8\textwidth]{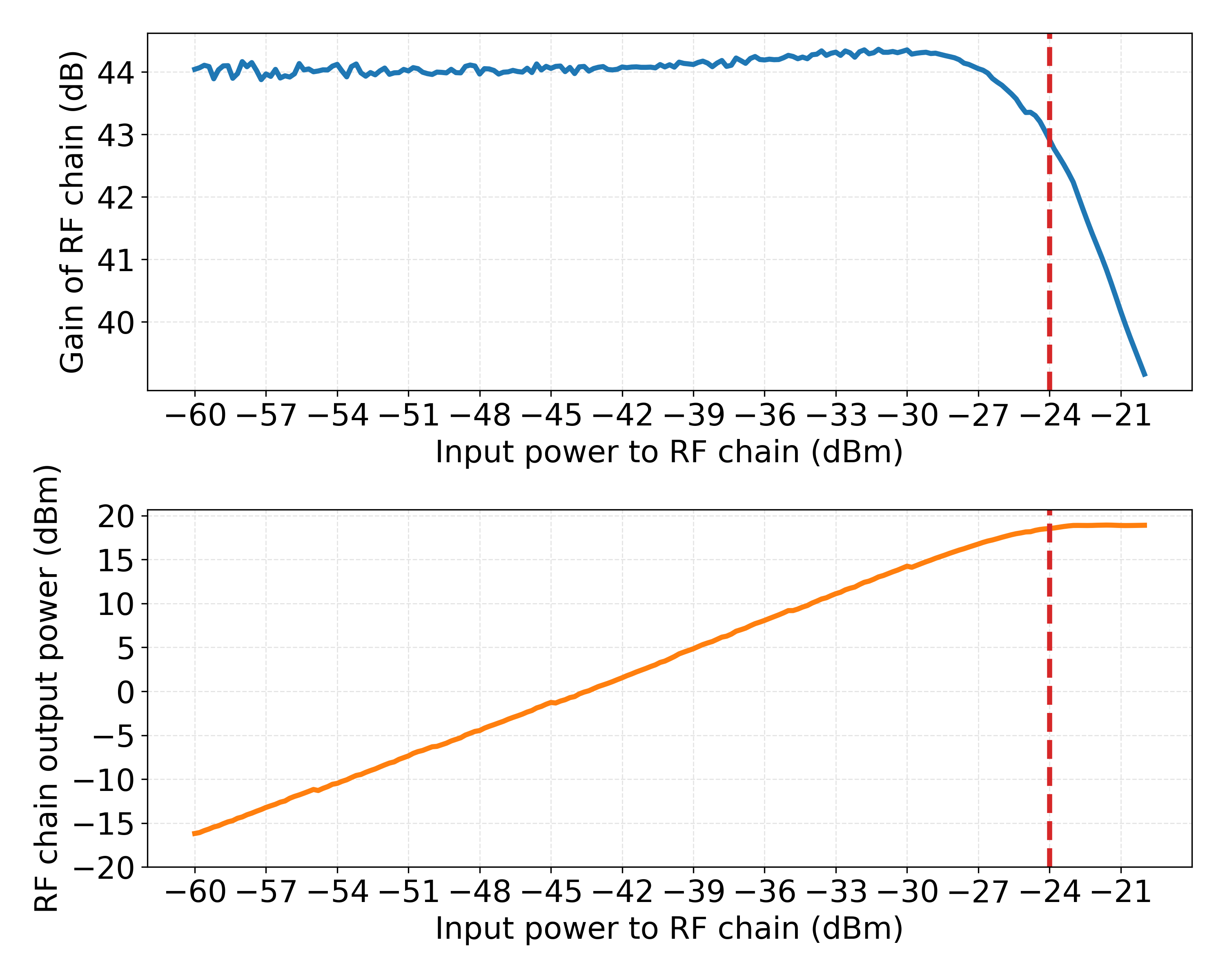} 
    \caption{The top panel shows the gain variation of the RF signal chain as a function of input power at the pre-amplifier input port. The bottom panel depicts the linearity plot of the RF signal chain. As the input power to the pre-amplifier module increases, the receiver chain moves from a linear regime to the 1~dB compression point of the amplifier in the post-amplifier unit. Both plots were obtained by sweeping the power of a tone at 150~MHz.}
    \label{fig:S21}
\end{figure}

The RF receiver chain comprises front-end and back-end electronics interconnected through a 10~m low-loss LMR 400 coaxial cable. The front-end electronics consists of a switchable filter unit and a pre-amplifier unit that amplifies the weak analog signal at the antenna output by $\sim$ 24~dB through an amplifier. A bias tee isolates the 24~V direct current from the RF signal path to provide DC bias to the RF amplifier. The filter section can either include a band-limiting filter in the RF path or bypass all filters to process the entire 30-300~MHz signal. Including an appropriate band-limiting filter helps minimize the effect of strong out-of-band interfering signals on the dynamic range of the RF analyzer. The back-end electronics consist of a post-amplifier unit with approximately 24~dB gain, an RF analyzer that produces the spectrum of the amplified signal, and a laptop. The laptop configures the settings\cite{Rauscher_Spectrum_Analysis} of the RF analyzer, such as the frequency span, resolution bandwidth, signal averaging, switching ON-OFF its pre-amplifier, and reference level, and acquires data via the Gigabit Ethernet link. The 24~V DC from the batteries powers the post-amplifier module. In the post-amplifier module, a bias tee carries 24~V DC into the RF signal path to power the pre-amplifier in the front-end electronics section. In the 30-300~MHz frequency range, the RF signal chain provides an effective gain of about 44~dB, including an insertion loss of 0.6~dB due to the 10~m long coaxial cable. The top panel of Figure \ref{fig:S21} shows the gain performance of the entire RF signal chain as a function of input power. It maintains an effective gain of about 44~dB as long as the signal chain operates linearly. To ensure linear operation during the surveys, the input power is kept below 1~dB compression point, which is the input power level at which gain reduces by 1~dB from its linear expectation, using attenuators or filters as needed. The bottom panel of Figure \ref{fig:S21} depicts the RF signal chain's input versus output power variation. As the input power to the RF signal chain increases beyond -30~dBm, the receiver chain begins to compress, reaching the 1~dB compression point for an input power of about -24~dBm. The input compression point sets the maximum power at the input of the front-end RF amplifier, usually dominated by RFI, detectable with this system.

\subsection{RF shielding and back-end Electronics}

The back-end electronics modules are housed inside an RF-shielded enclosure to shield the antenna from picking up the self-generated radio-frequency interference emanating from the laptop and RF Analyzer.  In the 30-300~MHz frequency range, 75~dB of average RF isolation provided by the shielding enclosure ensures the strength of the leaking signals is below the RF Analyzer’s noise floor. Additionally, a spacing of about 10~m between the antenna and the shielded enclosure will balance the need to minimize cable insertion loss and the coupling of leaking RFI from the laptop and RF Analyzer into the signal chain via the discone antenna. The data is recorded at a time resolution of $\sim$2~sec and a frequency resolution of $\sim$500~kHz. A continuous RFI data recording session lasts $\sim$4~hours, limited only by the battery charge level of the RF Analyzer.

\section{Observations} \label{sec: Observations}

The RFI survey was carried out at four remote locations that exhibited varying morphologies of RFI.
The observation duration at each site varied, typically spanning several hours, to capture RFI from sources that appear or disappear at different times throughout the day. 

The surveyed locations cover diverse terrains, including the Twin Lakes in Ladakh (Trans-Himalayan region), the Andaman Islands, Gruvebadet Atmosphere Laboratory in Ny-Ålesund, Svalbard, Norway, and the Gauribidanur Radio Observatory, Karnataka, India. The Gauribidanur site serves as a reference for understanding RFI conditions in locations closer to cities. An additional day scan was done at Twin Lakes, Ladakh, to compare the levels of RFI change between day and night.

\begin{table}[h!]
\centering
\caption{Site topological details for the observations, including geographic coordinates, horizon obstructions, and blockage solid angle.}
\label{table:site_details}

\begin{tabularx}{\textwidth}{|p{0.25\textwidth}|p{0.2\textwidth}|p{0.13\textwidth}|p{0.12\textwidth}|p{0.13\textwidth}|}
\hline
\textbf{Site} & \textbf{Coordinates} & \textbf{Mean Horizon (degrees)} & \textbf{Peak Horizon (degrees)} & \textbf{Blockage Solid Angle ($\pi$ units)} \\ \hline
Gauribidanur Radio Observatory, Karnataka, India (GRO) & 13.6° N, 77.4° E & 0.47 ± 0.32 & 1.14 & 0.0001\\ \hline
Twin Lakes, Ladakh, India (TLL) & 32.97° N, 78.58° E & 4.1 ± 2.21 & 10.9 & 0.0068\\ \hline
Kalpong Dam, Andaman Islands, India\ (KDA) & 13.1° N, 92.9° E & 5.7 ± 2.9 & 12.5 & 0.0126\\ \hline
Gruvebadet Atmosphere Laboratory, Ny-Ålesund, Svalbard, Norway (GLS) & 78.9° N, 11.9° E & 4.85 ± 4.76 & 17.8 & 0.0141 \\ \hline
\end{tabularx}

\end{table}

\begin{figure}[H]
    \centering
    \includegraphics[width=0.8\textwidth]{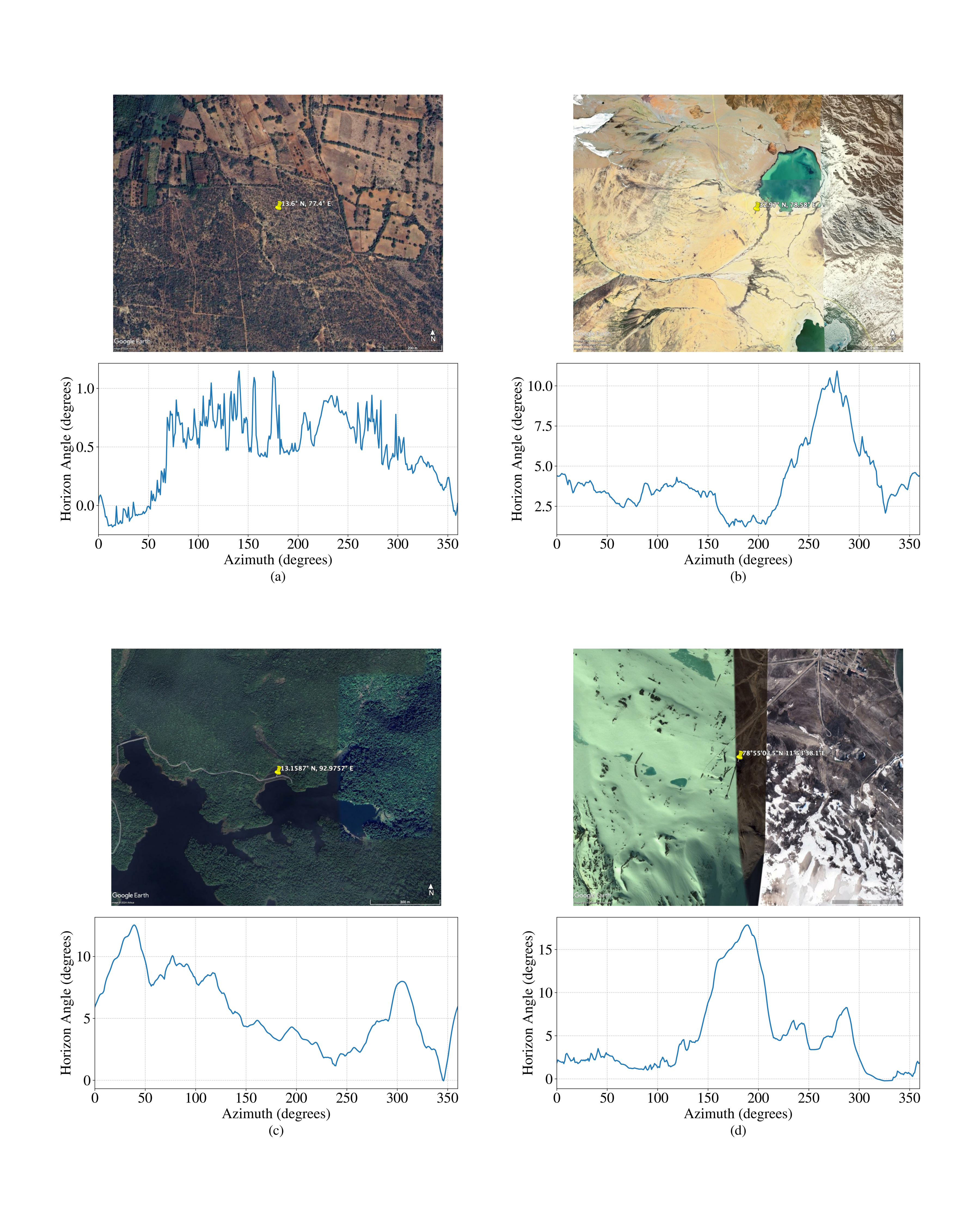}
    \caption{Satellite images of the surveyed locations along with the horizon profile around it obtained using the Python package \texttt{shapes}. (a) GRO (b) TTL (c) KDA (d) GLS.}
    \label{fig:terrain plots}
\end{figure}

In addition to the RFI environment, which is a primary criteria in selecting a site, other parameters such as horizon profile and accessible foregrounds are also considered in this study.
The terrain of these sites plays a critical role in shaping their RFI environments. Figure \ref{fig:terrain plots} shows the terrain and horizon plots for each surveyed location\cite{2021ApJ...923...33B}. The horizon is plotted using the Python package \texttt{shapes}\footnote{\href{https://github.com/npbassett/shapes}{https://github.com/npbassett/shapes}}. The horizon is particularly influential in the site's exposure to RFI. Locations with significant obstructions at the horizon can block terrestrial RFI from those directions, providing natural shielding\cite{thompson2017interferometry}. However, while a high horizon may shield against terrestrial RFI, it complicates foreground modeling if the horizon lies in the line of sight of the beam response of the antenna and obstructs the sky. This becomes critical during the modeling of scientific data\cite{2024MNRAS.527.2413P}. This issue is particularly relevant to SARAS-like experiments where the maximum response is directed towards the horizon. 

Among the sites, the GRO had the least obstruction at the horizon, with a mean obstruction of 0.47 degrees, a maximum of 1.14~degrees, and a standard deviation of 0.32~degrees along the azimuth. Table \ref{table:site_details} gives details of the mean obstruction at the horizon and the maxima of the profile. The detailed horizon profiles are shown in Figure \ref{fig:terrain plots} along with the satellite view of each site.

For this study, although the sky varies across the surveyed locations and times of observation, the antenna's broad beam averages over a large portion of the sky. This spatial averaging dilutes the foreground structures, rendering them smooth with no sharp variations along the frequency axis\cite{2008MNRAS.388..247D}\cite{2017AJ....153...26S}. Foreground emissions from galactic and extragalactic sources change gradually across the sky, exhibiting continuity across drift scans. In contrast, human-generated RFI, in general, manifests as sharp, discrete features in specific, narrow frequency channels and at specific times of the day such as satellites, or broadcast communications. Additionally, some RFI may be persistent or broad-band in nature. The spectral characteristics and temporal occurrence of such RFI will be discussed in detail in later sections of this paper.

\begin{figure}[H]
    \centering
    \includegraphics[width=\textwidth]{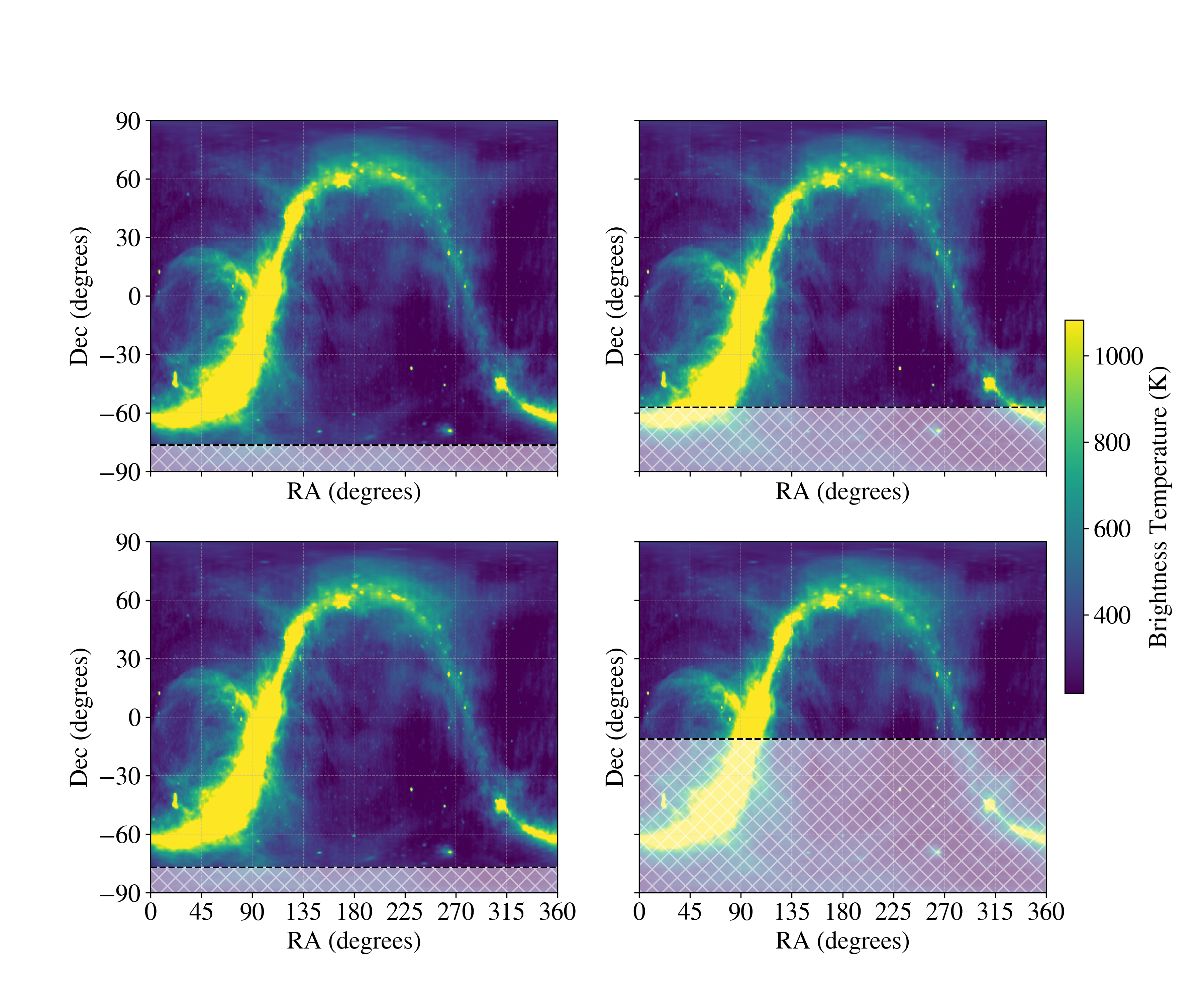}
    \caption{Observable foregrounds (unshaded region) at each site, GRO (top-left), TLL (top-right), KDA (bottom-left) and GLS (bottom-right). The maps are generated at a frequency of 140~MHz using the Global Sky Model (GSM).}
    \label{fig: foregrounds}
\end{figure}

For science deployment, understanding the sky at each site plays a key role. All of our surveyed sites are in the northern hemisphere. This means that a certain degree of the southern sky can never be observed from these sites at any time of the year. Figure \ref{fig: foregrounds} shows the observable sky at each location as predicted by the Global Sky Model at 140~MHz\cite{2008MNRAS.388..247D}. The shaded region in each subplot is the respective non-observable sky at these locations. A combination of terrain profiles and accessibility of foregrounds from each site will be crucial to developing a robust model for the foreground\cite{761c98bcdd13467f87dd40e7180307b3}.

\subsection{System calibration}

\begin{figure}[H]
    \centering
    \begin{subfigure}{0.50\textwidth}
        \centering
        \includegraphics[width=\linewidth]{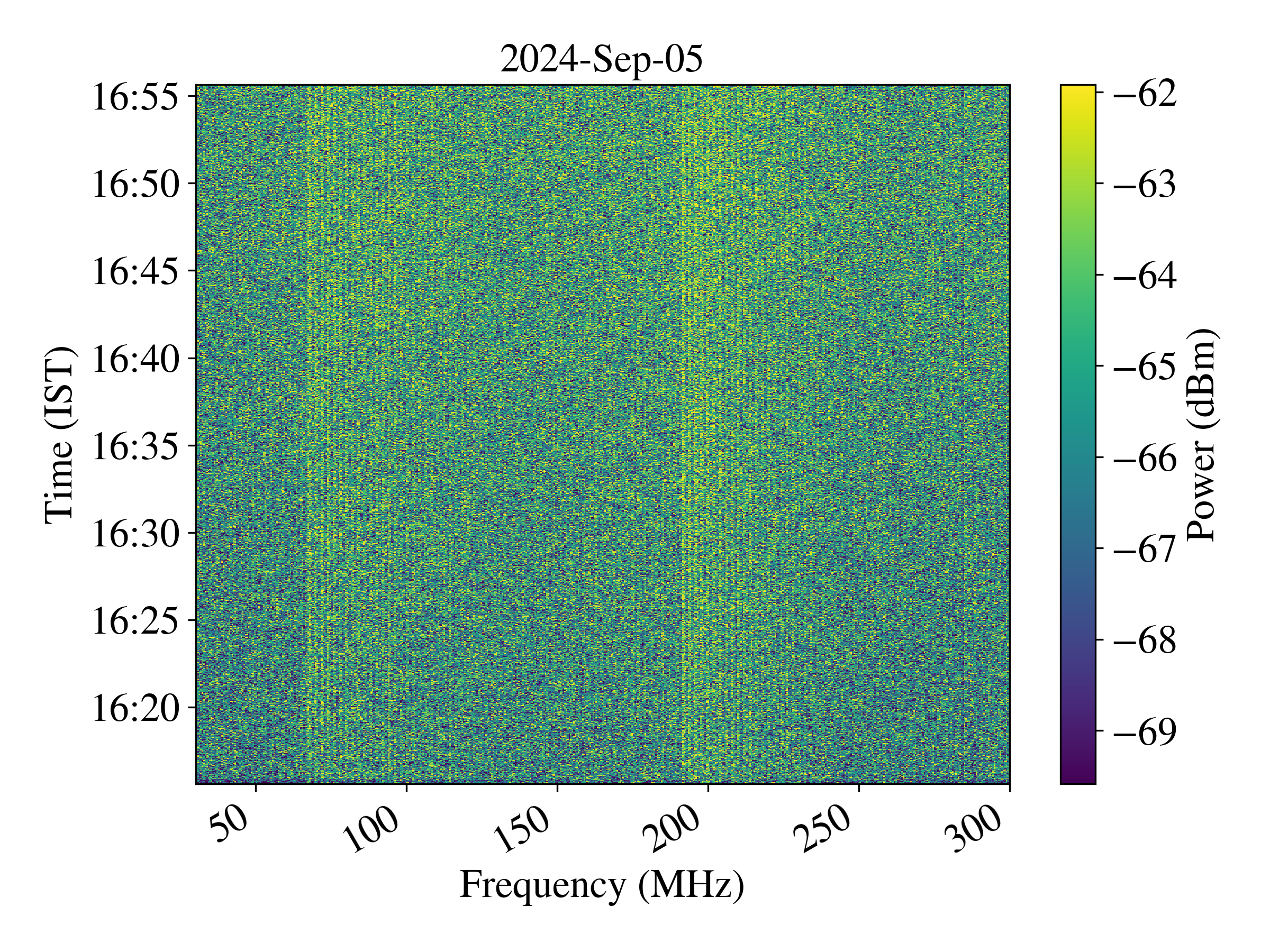}
        \caption{}
        \label{subfig:5-500_load_waterfall}
    \end{subfigure}
    \hfill
    \begin{subfigure}{0.5\textwidth}
        \centering
        \includegraphics[width=\linewidth]{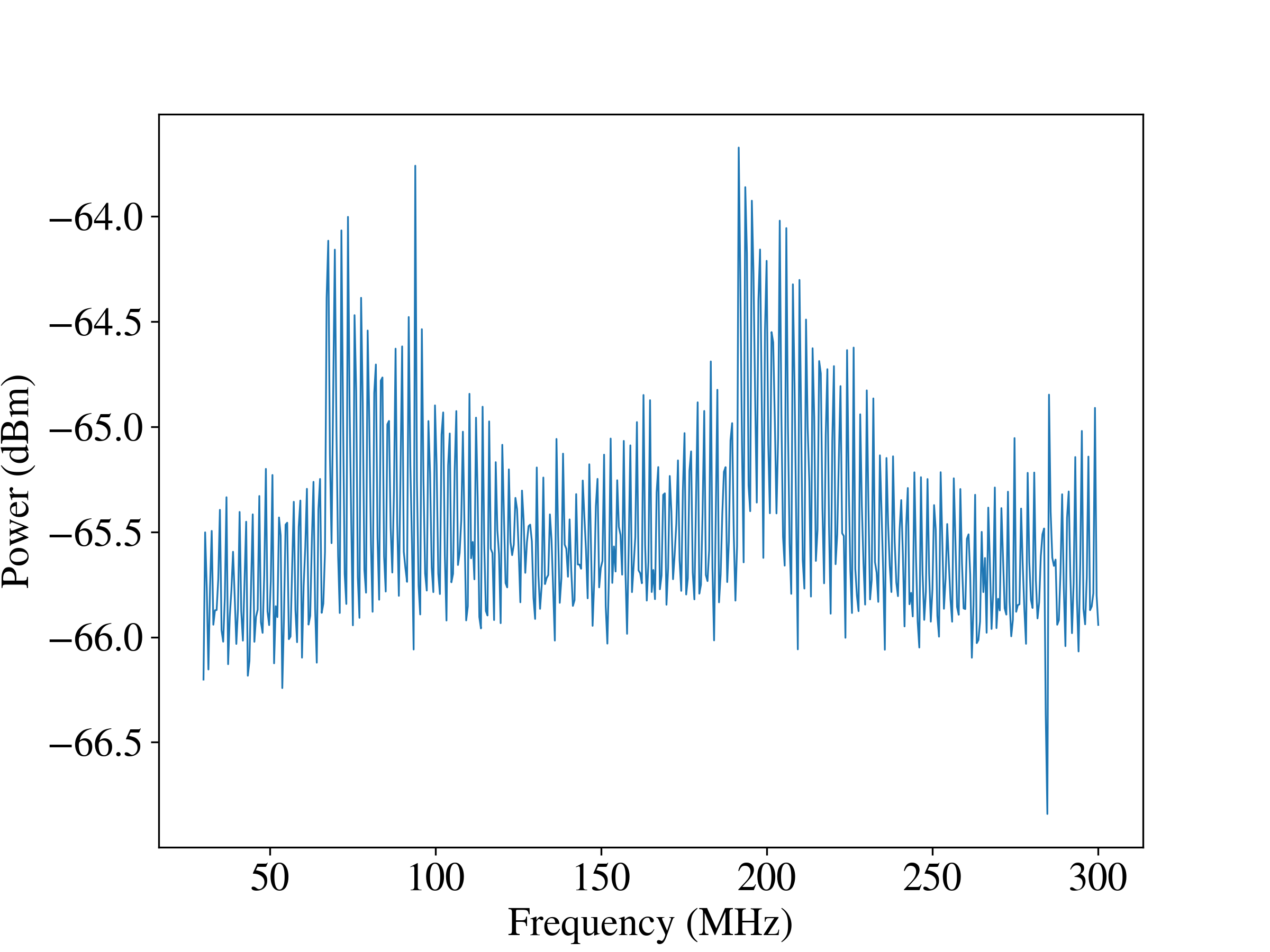}
        \caption{}
        \label{subfig:5-500_load_median}
    \end{subfigure}
    \caption{System gain features RFI radiometer. (a) drift scan of the 50-ohm load termination, (b) median of the drift scan along the temporal axis.}
    \label{fig:sidebyside 50 ohm}
\end{figure}

System calibration aims to correct for the multiplicative bandpass of the radiometer. The calibration process used a 50-ohm termination as a reference load to correct for frequency-dependent bandpass, such as amplifier gains and path losses, allowing accurate estimation of the sky spectrum. Measurements were taken with the 50-ohm load at GRO, replacing the antenna with a well-matched 50-ohm termination at the input of the pre-amplifier. The ambient temperature for calibration was 290~K. Figure \ref{fig:sidebyside 50 ohm} shows the drift scan of a 50-ohm termination. Figures \ref{subfig:5-500_load_waterfall} and \ref{subfig:5-500_load_median} show system bandpass, especially within the 65-120~MHz and 190-250~MHz sub-bands, where features particularly introduced due to the noise floor of the instrument are visibly evident.

Using a 50-ohm load at 290~K as a reference, a conversion factor was established to calibrate the measured data. Antenna and load measurements were first converted from dBm to Kelvin units using equation  (\ref{eq:watts to kelvin}),

\begin{equation}
    T_{\text{meas}}(K) = \frac{10^{\frac{P_{\text{{dBm}}}}{10}} \times 10^{-3}}{k_B \cdot \text{bandwidth}}
    \label{eq:watts to kelvin}
\end{equation}

where \( P_{\text{dBm}} \) represents measurements, either from the antenna or reference load, and \( k_B \) is the Boltzmann constant, and the bandwidth is set to 0.5~MHz. \( T_{\text{meas}}(K) \) represents either antenna \( T_{\text{antenna}}(K) \) or load \( T_{\text{Load}}(K) \) measurements converted to Kelvin.

The standard ambient temperature \( T_{\text{ambient}}(K) \) of 290~K was divided by the load measurements \( T_{\text{Load}}(K) \) to calculate a conversion factor,

\begin{equation}
    \text{Conversion Factor} = \frac{T_{\text{ambient}}(K)}{T_{\text{Load}}(K)}
    \label{eq:conversion factor}
\end{equation}

Finally, the factor was applied to the antenna measurement for bandpass calibration:

\begin{equation}
    T_{\text{calibrated}}(K) = T_{\text{antenna}}(K) \cdot \text{Conversion Factor}
    \label{eq:corrected temperature}
\end{equation}

This calibration provides a spectrum measurement free from system gain as a function of frequency. This calibrated measurement is then converted back to dBm units for further visualization due to the high dynamic range.

\begin{figure}[htb]
    \centering
    \begin{subfigure}{0.48\textwidth}
        \centering
        \includegraphics[width=\linewidth]{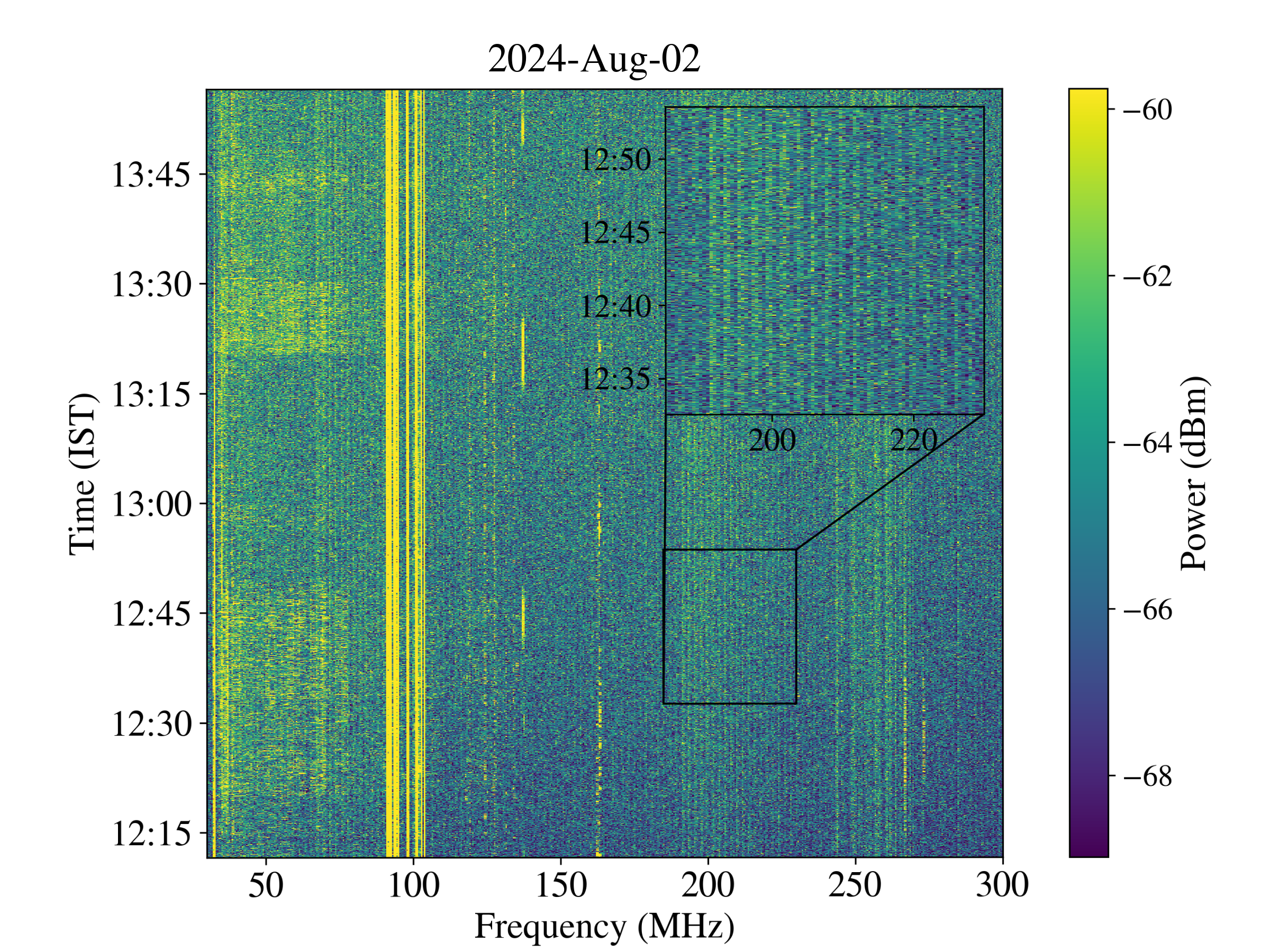}
        \caption{}
        \label{subfig:GBD_uncal_waterfall}
    \end{subfigure}
    \hfill
    \begin{subfigure}{0.48\textwidth}
        \centering
        \includegraphics[width=\linewidth]{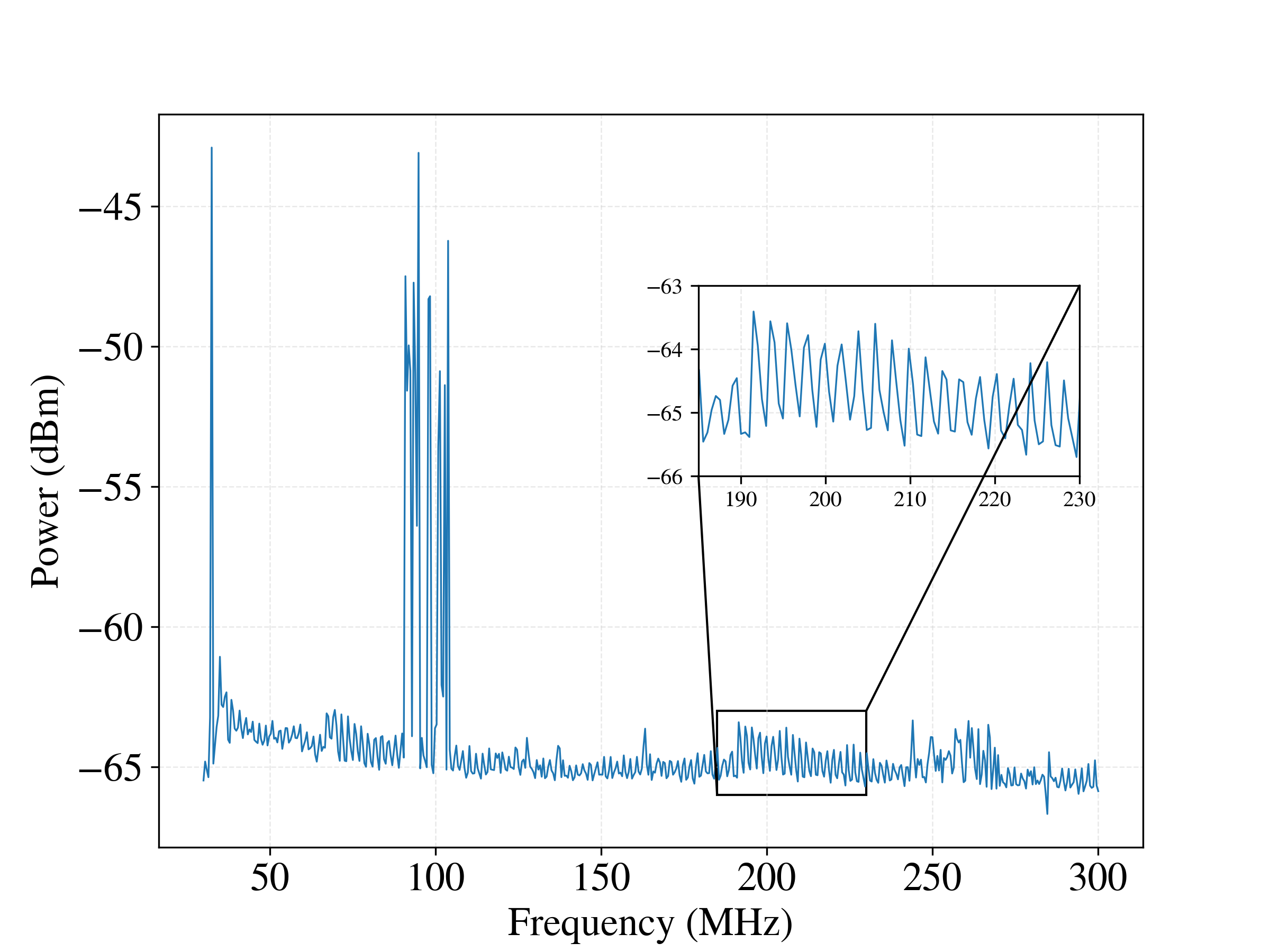}
        \caption{}
        \label{subfig:GBD_uncal_median}
    \end{subfigure}
    
    \vspace{0.4cm} 

    \begin{subfigure}{0.48\textwidth}
        \centering
        \includegraphics[width=\linewidth]{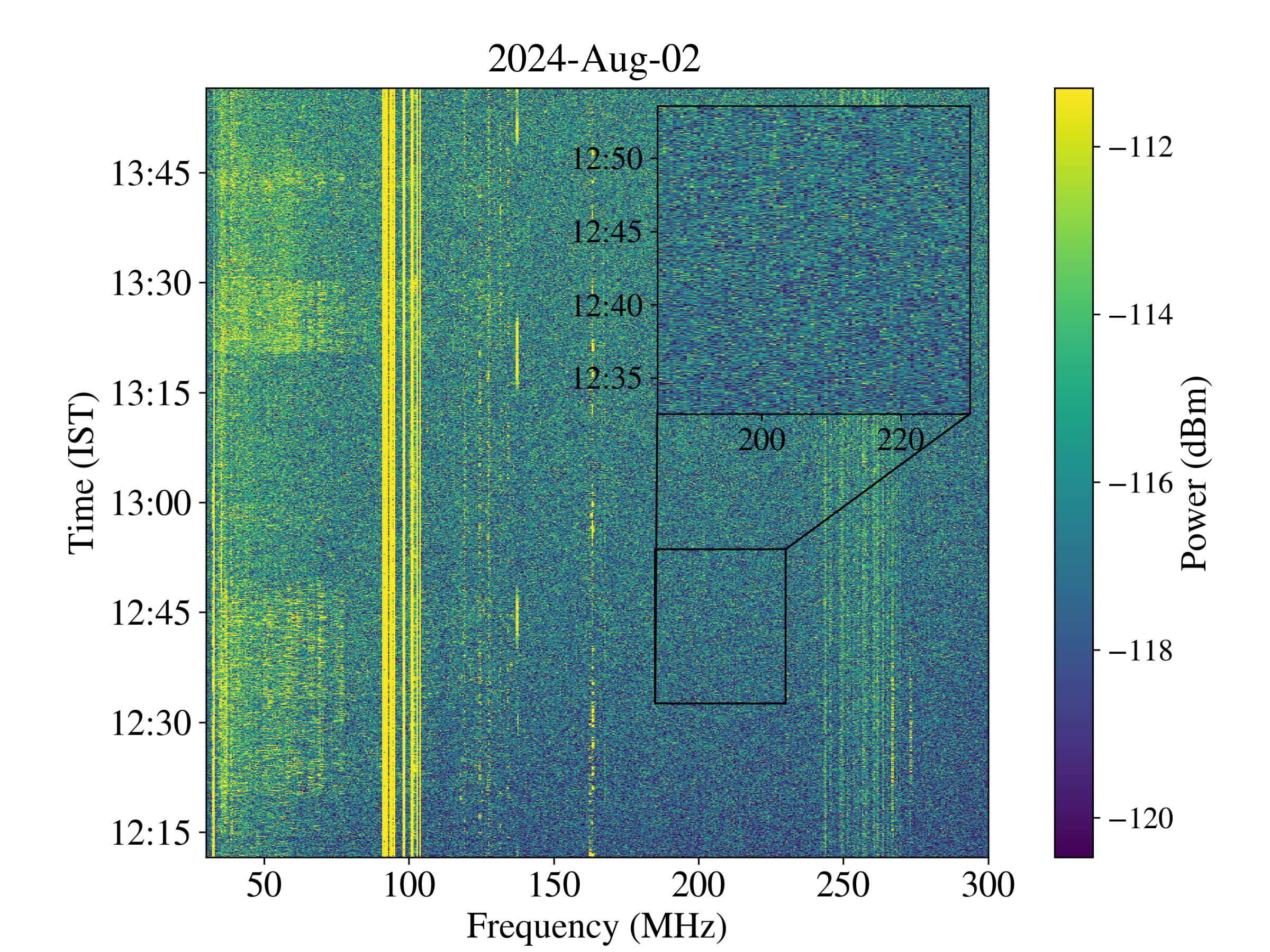}
        \caption{}
        \label{subfig:GBD_cal_waterfall}
    \end{subfigure}
    \hfill
    \begin{subfigure}{0.48\textwidth}
        \centering
        \includegraphics[width=\linewidth]{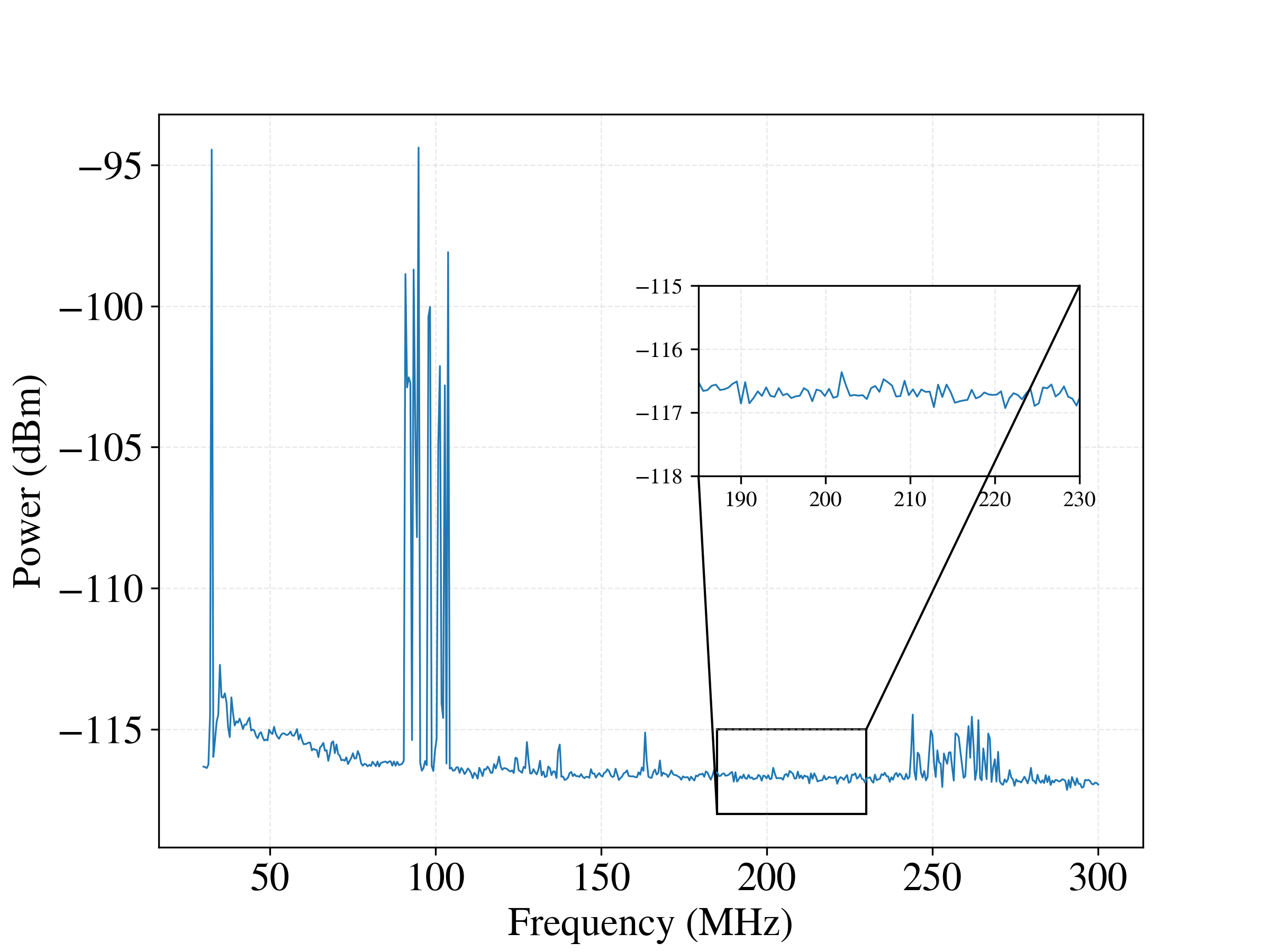}

        \caption{}
        \label{subfig:GBD_cal_median}
    \end{subfigure}

    \caption{(a) Drift scan of uncalibrated data from the Gauribidanur Radio Observatory, (b) temporal median of uncalibrated drift scan, (c) drift scan calibrated with a 50-ohm termination, and (d) temporal median of the calibrated drift scan.}
    \label{fig:Load_uncal_and_cal_GBD}
\end{figure}

Figure \ref{subfig:5-500_load_waterfall} and \ref{subfig:5-500_load_median} illustrate the system gain features that distort the measured sky spectrum. The gain features are clearly visible by connecting a 50-ohm termination replacing the antenna. The top panel of Figure \ref{fig:sidebyside 50 ohm} shows the power along frequency and time. A clearer spectral behaviour is observed in the bottom panel of Figure \ref{fig:sidebyside 50 ohm}, obtained by taking a median over time. 

Similar bandpass features were seen when the antenna was connected, as shown in Figure \ref{fig:Load_uncal_and_cal_GBD}. Figure \ref{subfig:GBD_uncal_waterfall} and \ref{subfig:GBD_uncal_median} show raw measurement and corresponding median spectrum, respectively, taken over $\sim$1.5 hours across 30-300~MHz. In addition to RFI, we distinctly see bandpass features, likely arising from the frequency-dependent noise floor of the RF analyzer. After applying bandpass calibration, these features were eliminated, as evident in Figures \ref{subfig:GBD_cal_waterfall} and \ref{subfig:GBD_cal_median}, where the data reaches the expected noise floor levels. While these examples are from GRO, the same corrections were also applied to data from KDA and GLS. Even though the ambient temperature varied at different surveyed locations, the spectral features in the system bandpass were found to remain consistent based on analyzing 50-ohm termination datasets from sites.

\begin{figure}[h!]
    \centering
    \includegraphics[width=1.1\textwidth]{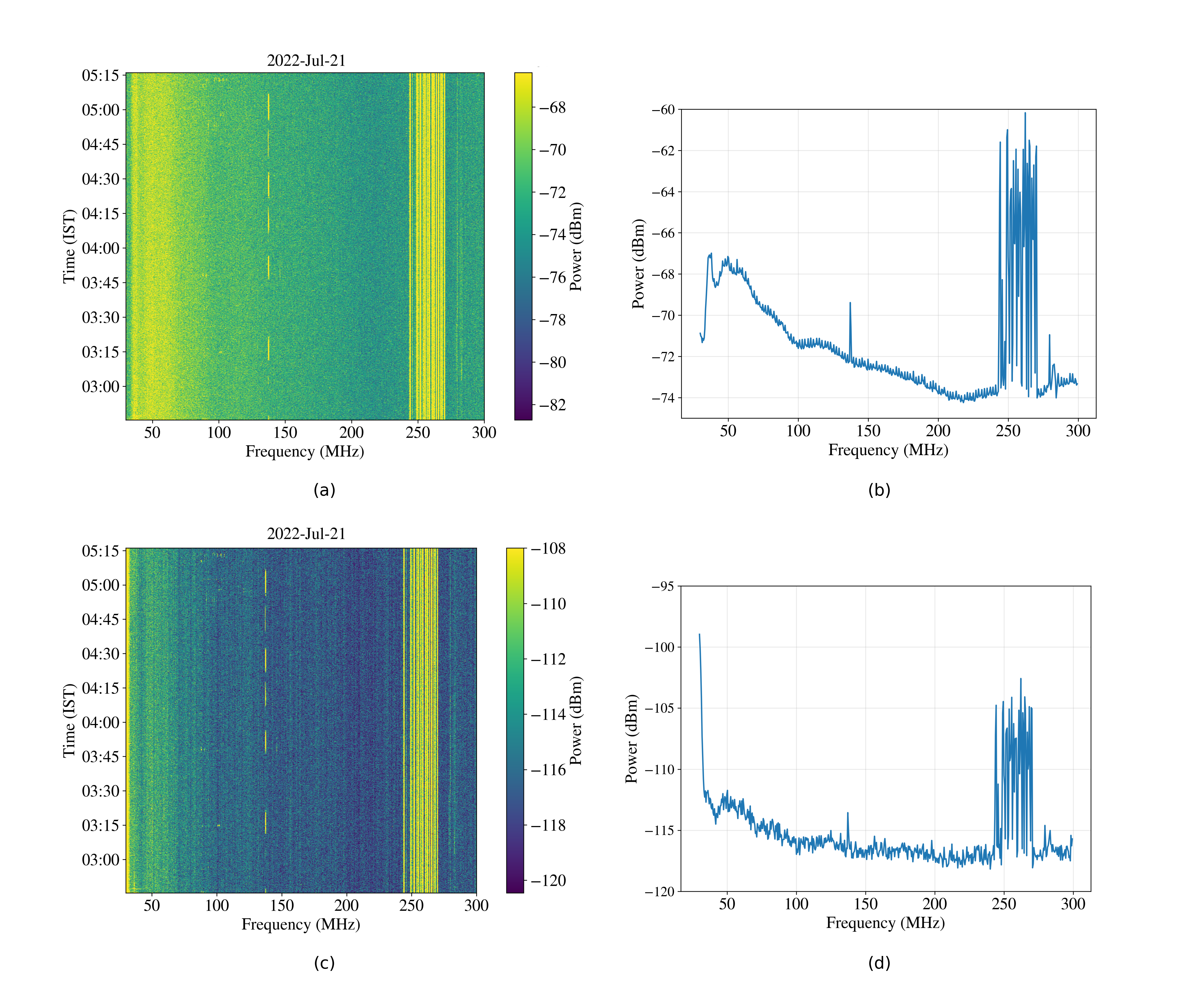}
    \caption{(a) Drift scan of uncalibrated data from the Twin Lakes Ladakh, (b) temporal median of uncalibrated drift scan, (c) calibrated drift scan, and (d) temporal median of the calibrated drift scan.}
    \label{fig:Ladakh_caibration}
\end{figure}

During the acquisition of data in TLL, a 30~MHz high-pass filter (HPF) was used to avoid saturation of the amplifier due to high out-of-band RFI at lower frequencies. The calibration of this data required a different approach. In this case, due to the high noise temperature of the amplifier ($\sim$400~K), bandpass could not account for the filter response. To address this, a separate measurement was conducted using a noise source with noise output $-95~dBm~Hz^{-1}$, which elevated the signal-to-noise ratio, ensuring that the filter's response is not buried under the amplifier’s noise. The gain measurement was used to calibrate the data from TLL. Figure \ref{fig:Ladakh_caibration} (a) and (b) show the uncalibrated drift scan and its median along the temporal axis from TLL, while Figure \ref{fig:Ladakh_caibration} (b) and (d) show the drift scan and its median after calibration.
A similar calibration method as used earlier was applied using these noise source measurements. Our subsequent visualizations and relevant analysis of the RFI environment will use these calibrated measurements, enabling a robust estimate of RFI statistics and their morphologies.

\subsection{Drift Scans}
\begin{figure}[htb]
    \centering
   
    \begin{subfigure}{0.5\textwidth}
        \centering
        \includegraphics[width=\linewidth]{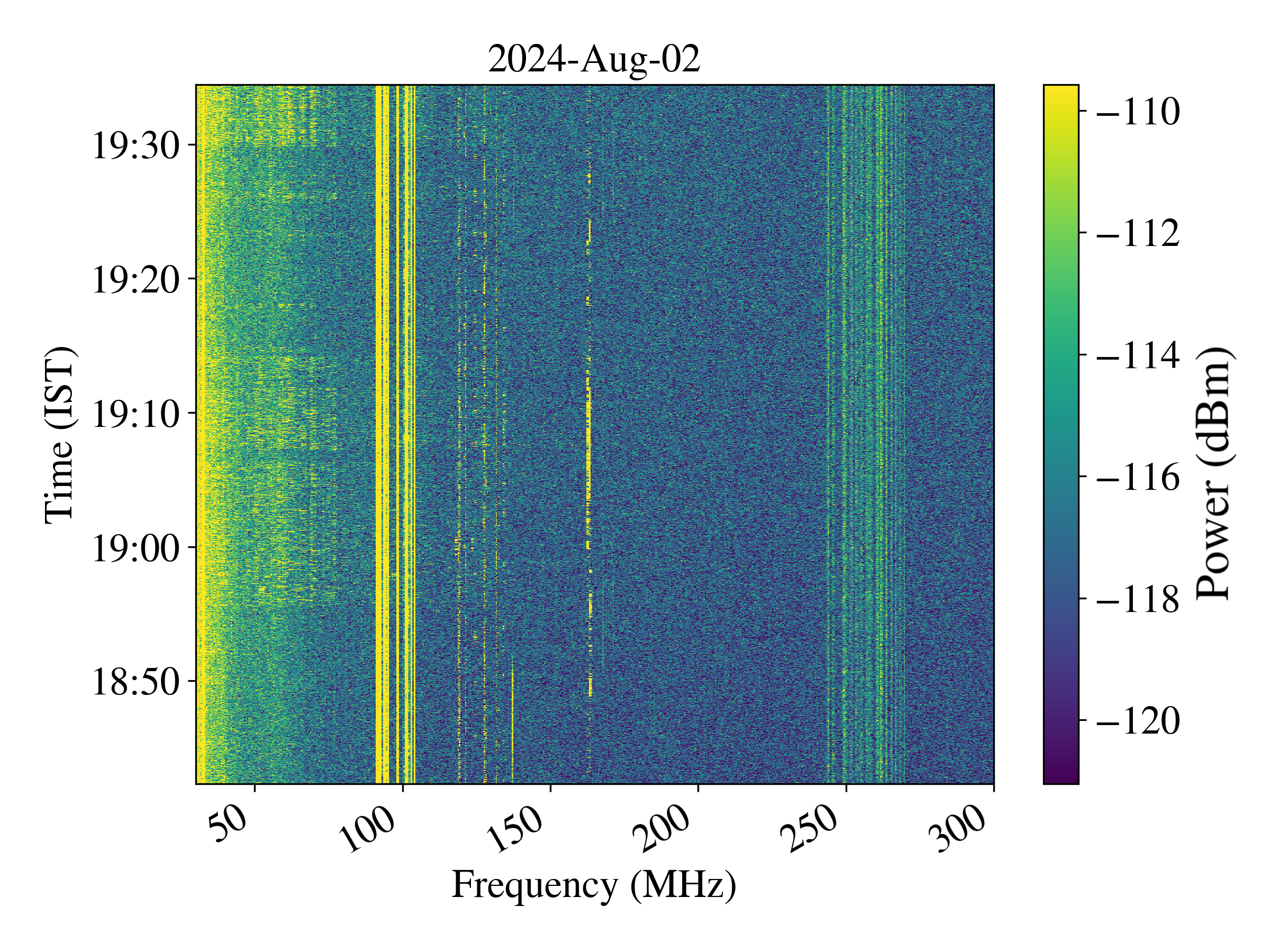}
        \caption{}
        \label{subfig:GBD_waterfall}
    \end{subfigure}
    \hfill
    \begin{subfigure}{0.45\textwidth}
        \centering
        \includegraphics[width=\linewidth]{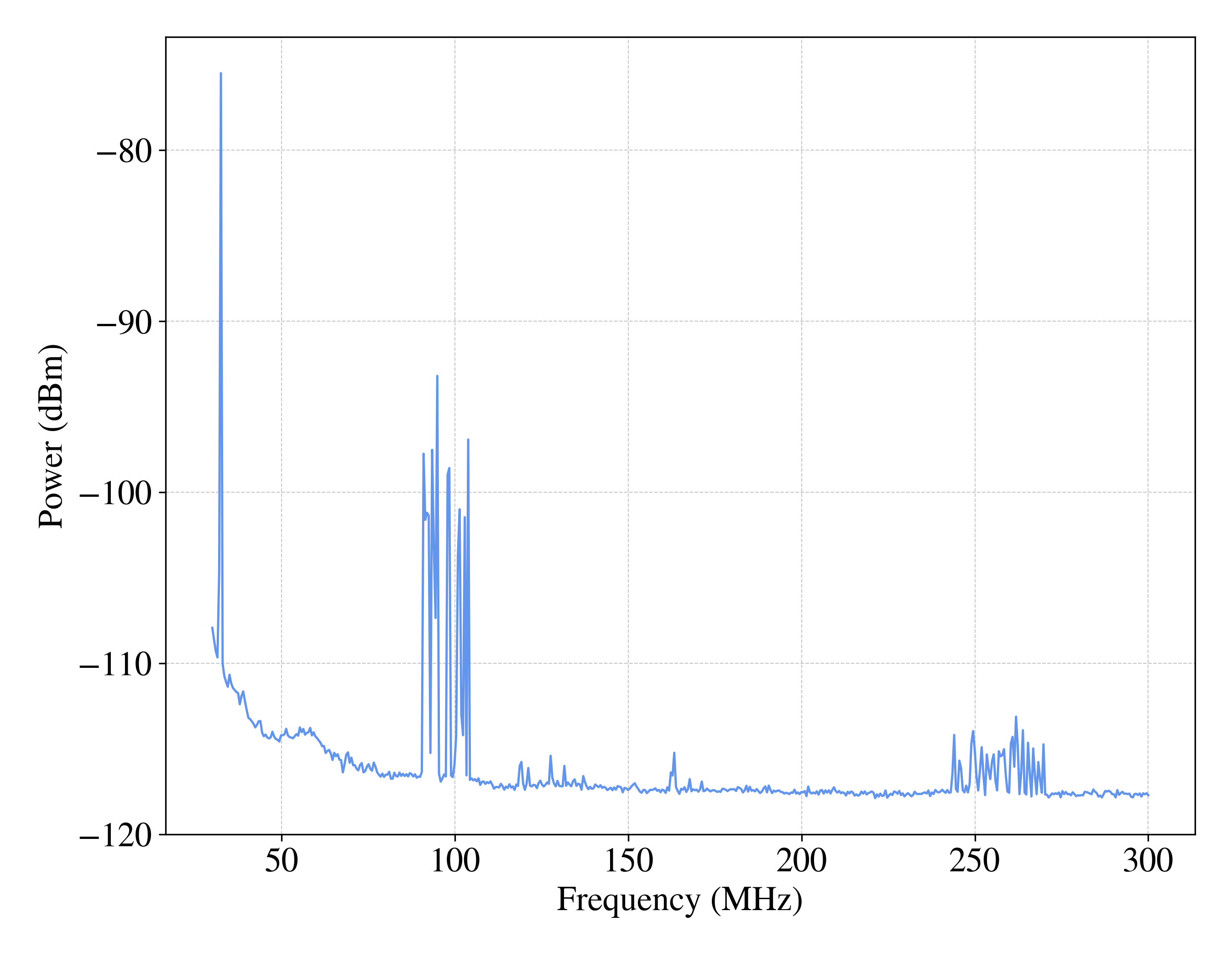}
        \caption{}
        \label{subfig:GBD_median}
    \end{subfigure}
    \caption{GRO drift scan highlighting RFI characteristics in the 30–300~MHz band. (a) Calibrated drift scan and (b) median along the temporal axis.}

    \label{fig:GBD calibrated waterfalls}
\end{figure}

\begin{figure}[htb]
    \centering

    \begin{subfigure}{0.5\textwidth}
        \centering
        \includegraphics[width=\linewidth]{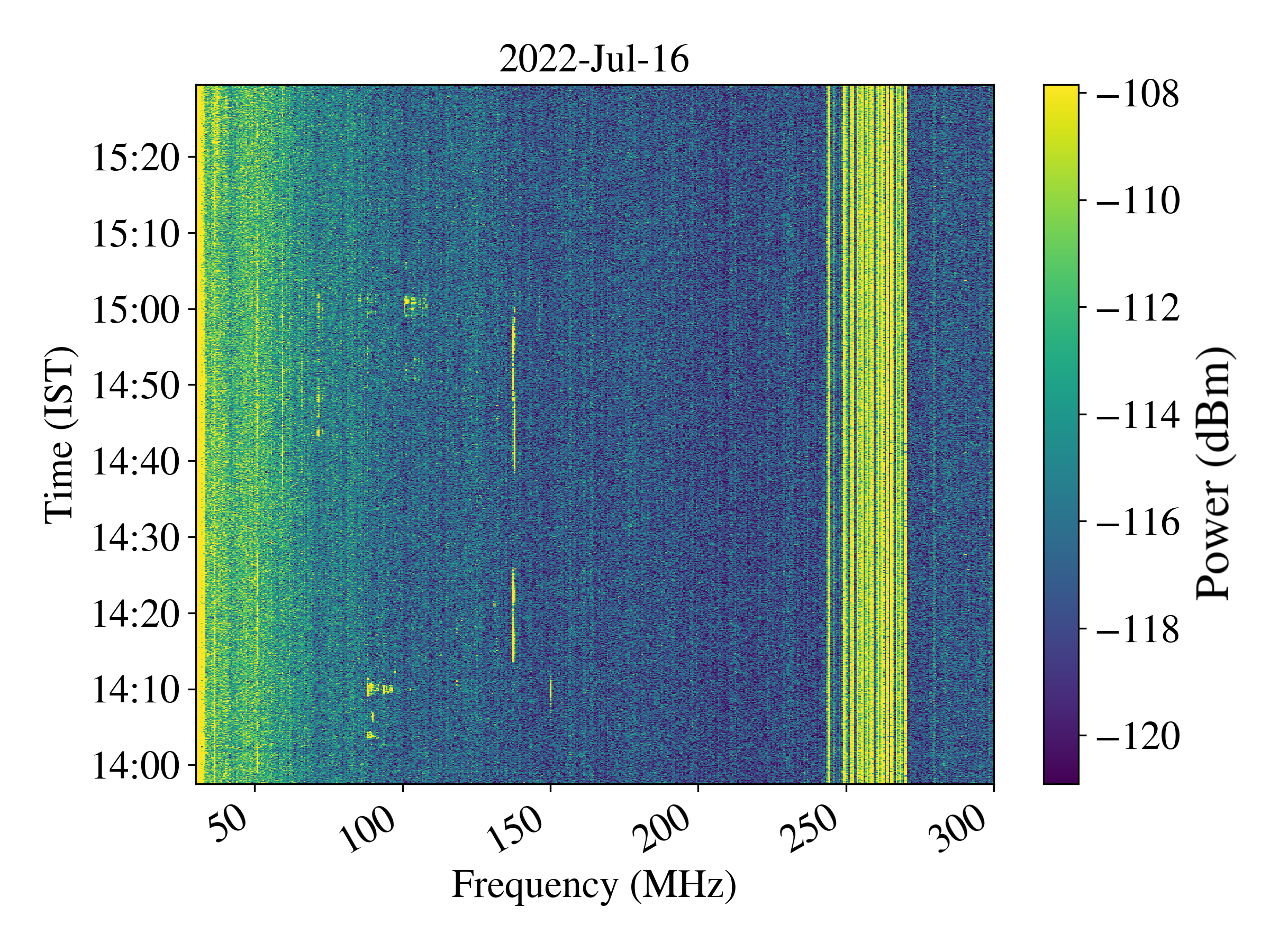}
        \caption{}
        \label{subfig:TLL_day_waterfall}
    \end{subfigure}
    \hfill
    \begin{subfigure}{0.45\textwidth}
        \centering
        \includegraphics[width=\linewidth]{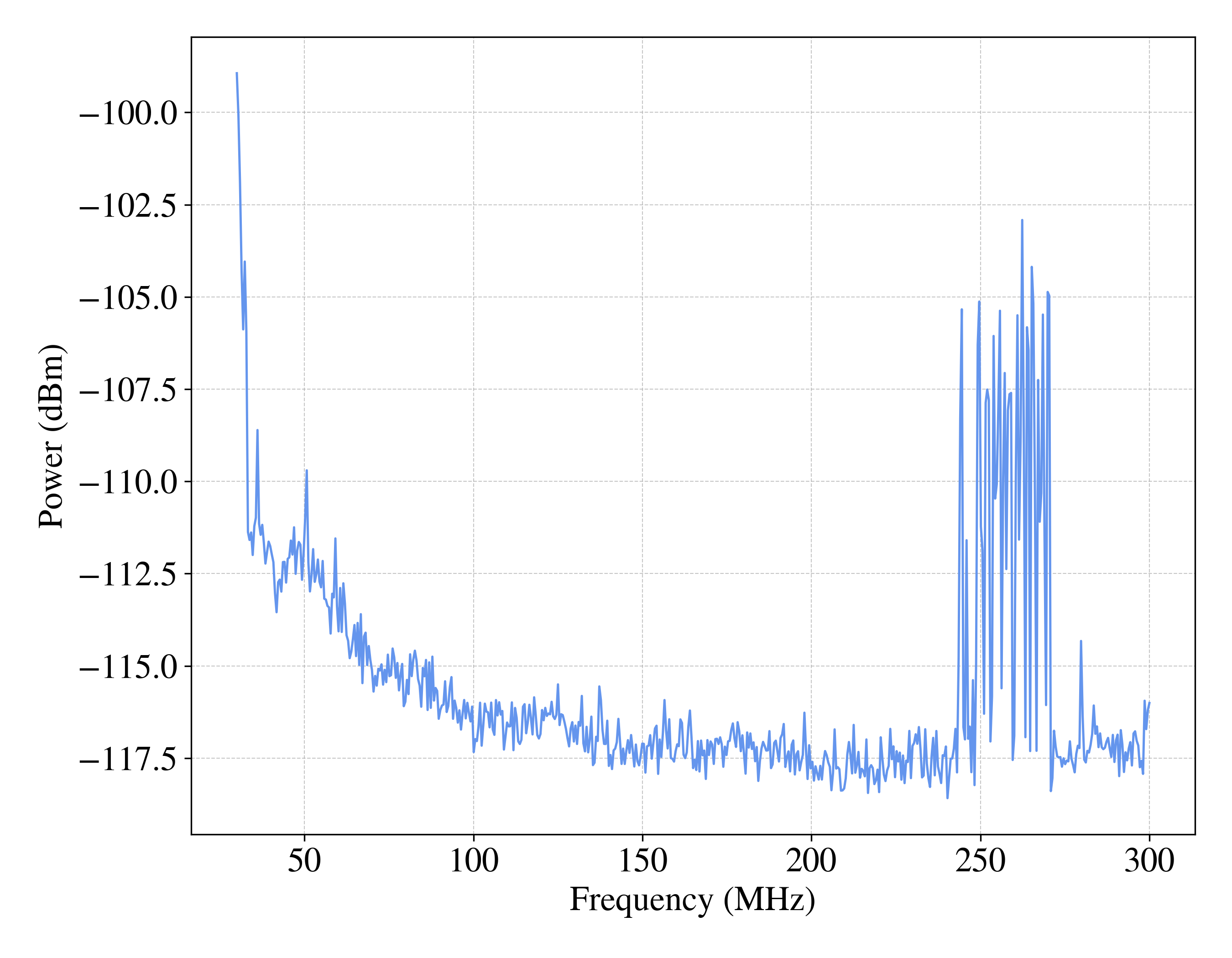}
        \caption{}
        \label{subfig:TLL_day_median}
    \end{subfigure}
    
    \vspace{10pt}

    \begin{subfigure}{0.5\textwidth}
        \centering
        \includegraphics[width=\linewidth]{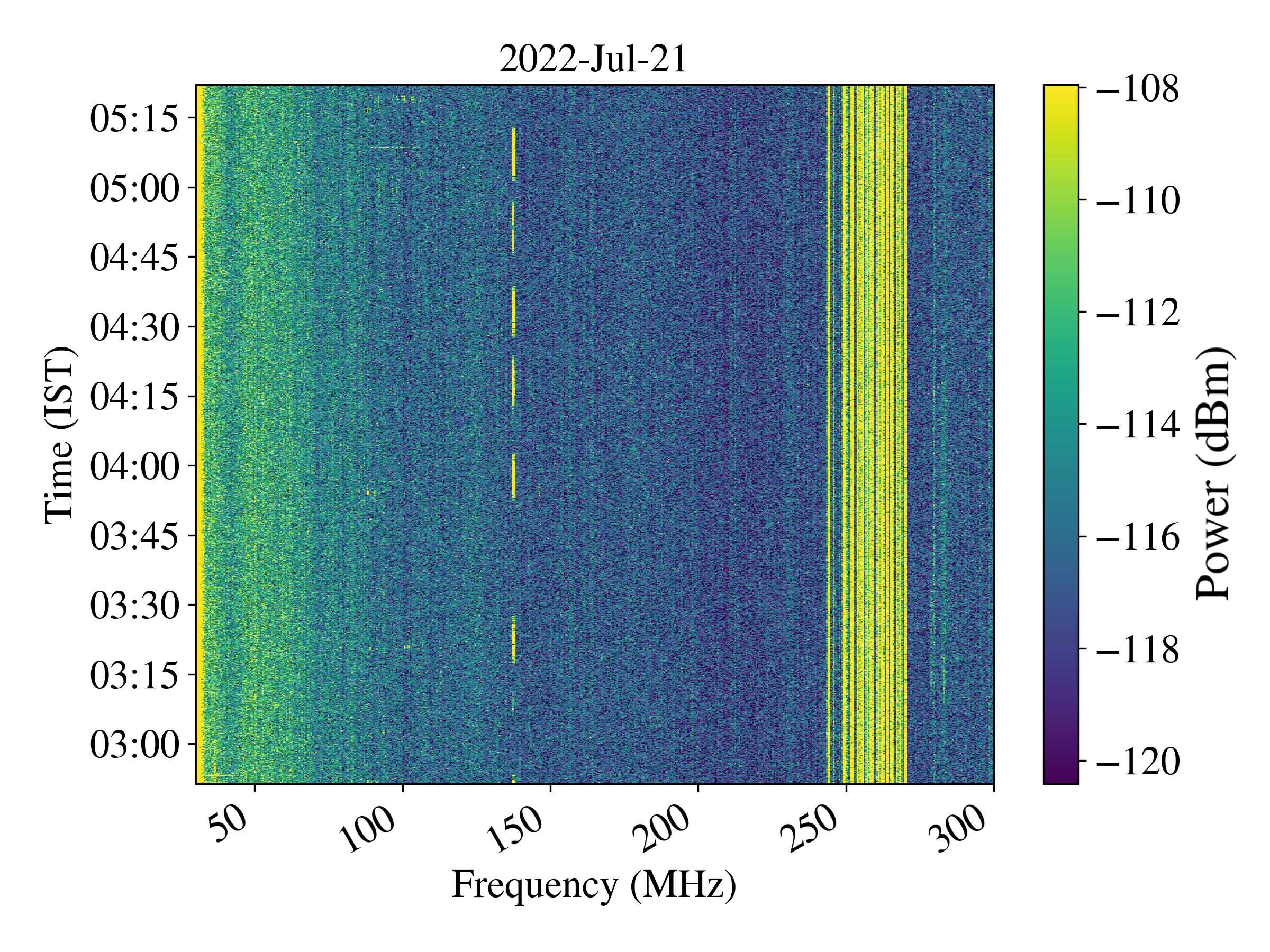}
        \caption{}
        \label{subfig:TLL_night_waterfall}
    \end{subfigure}
    \hfill
    \begin{subfigure}{0.45\textwidth}
        \centering
        \includegraphics[width=\linewidth]{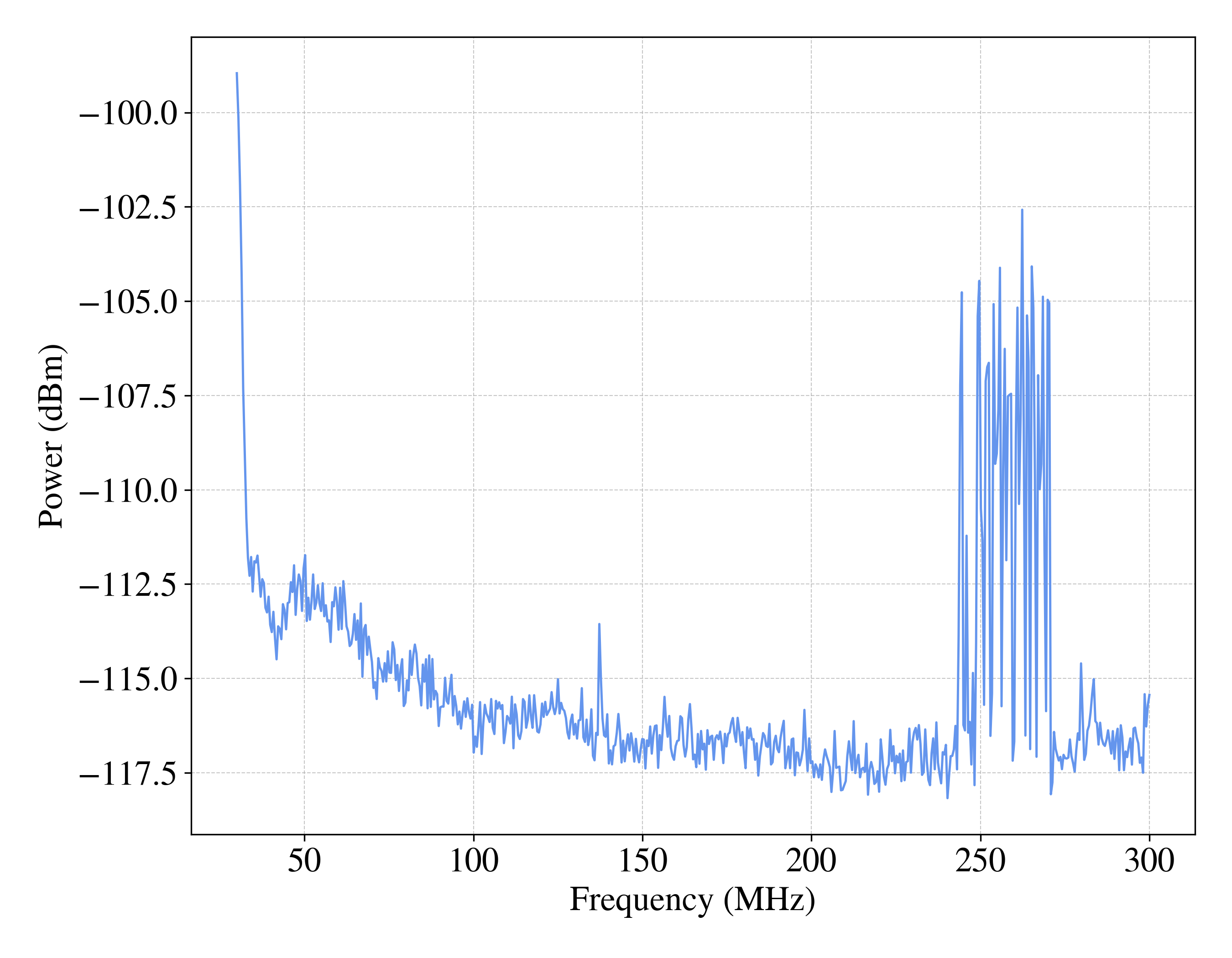}
        \caption{}
        \label{subfig:TLL_night_median}
    \end{subfigure}
    
    \caption{TLL drift scans highlighting RFI characteristics in the 30–300~MHz band. (a, c) Drift scans showing temporal variations for day and night, respectively, and (b, d) median along the temporal axis for the corresponding drift scans.}

    \label{fig:Ladakh calibrated waterfalls}
\end{figure}

\begin{figure}[htb]
    \centering
   
    \begin{subfigure}{0.5\textwidth}
        \centering
        \includegraphics[width=\linewidth]{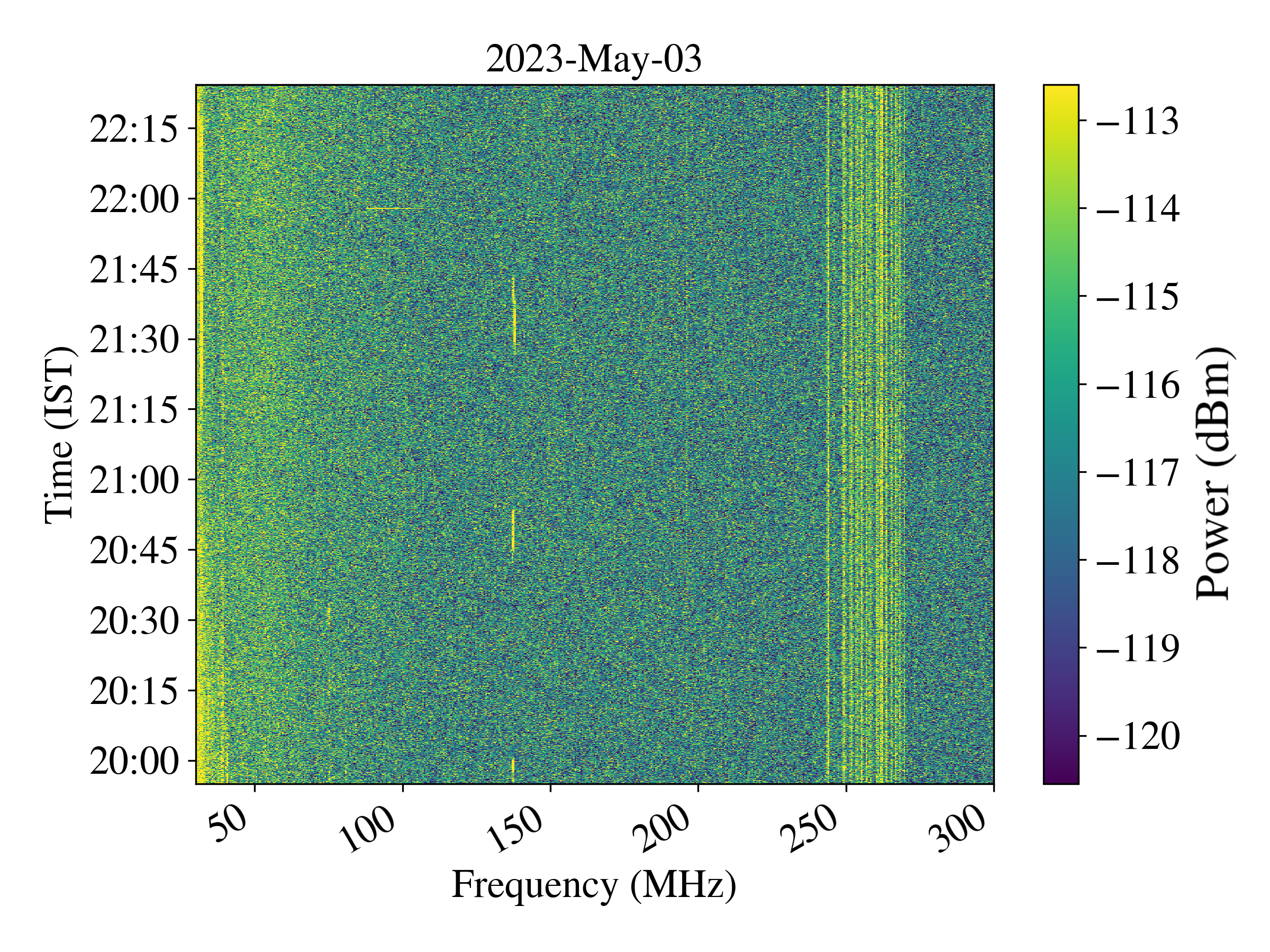}
        \caption{}
        \label{subfig:kalpong_waterfall}
    \end{subfigure}
    \hfill
    \begin{subfigure}{0.45\textwidth}
        \centering
        \includegraphics[width=\linewidth]{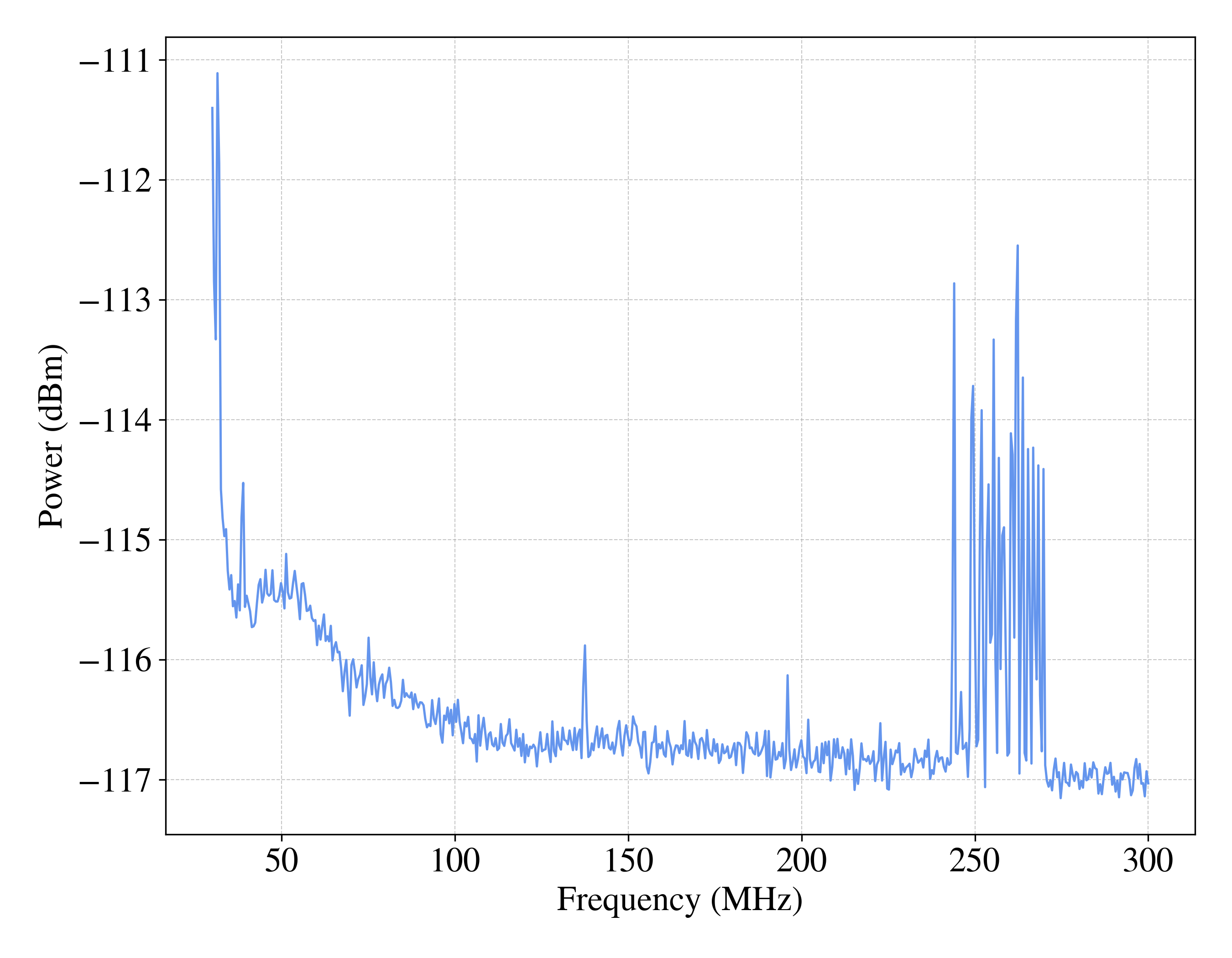}
        \caption{}
        \label{subfig:kalpong_median}
    \end{subfigure}
    \caption{KDA drift scan highlighting RFI characteristics in the 30–300~MHz band. (a) Calibrated drift scan and (b) median along the temporal axis.}

    \label{fig:Kalpong calibrated waterfalls}
\end{figure}

\begin{figure}[htb]
    \centering
    \begin{subfigure}{0.5\textwidth}
        \centering
        \includegraphics[width=\linewidth]{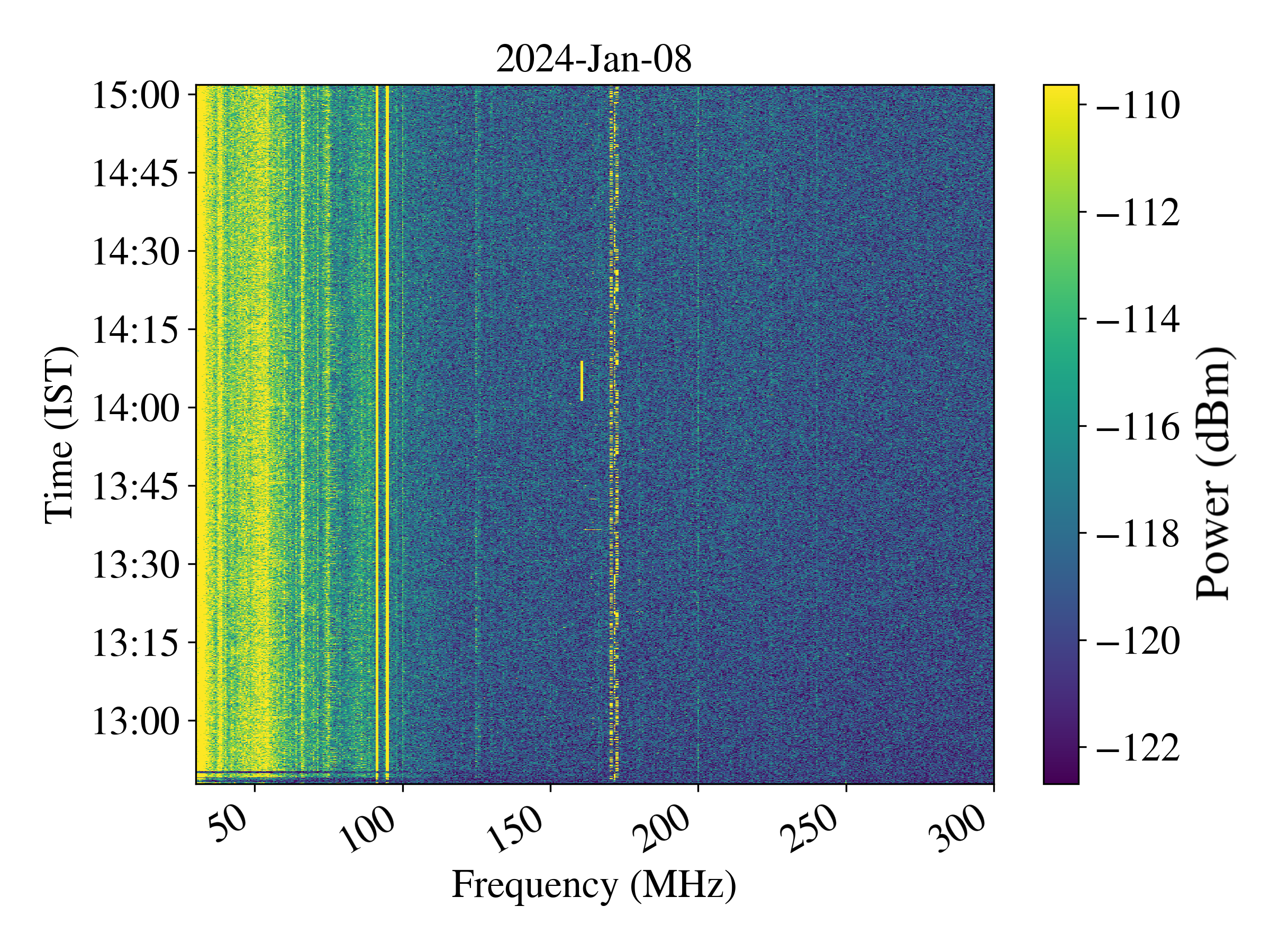}
        \caption{}
        \label{subfig:arctic_waterfall}
    \end{subfigure}
    \hfill
    \begin{subfigure}{0.45\textwidth}
        \centering
        \includegraphics[width=\linewidth]{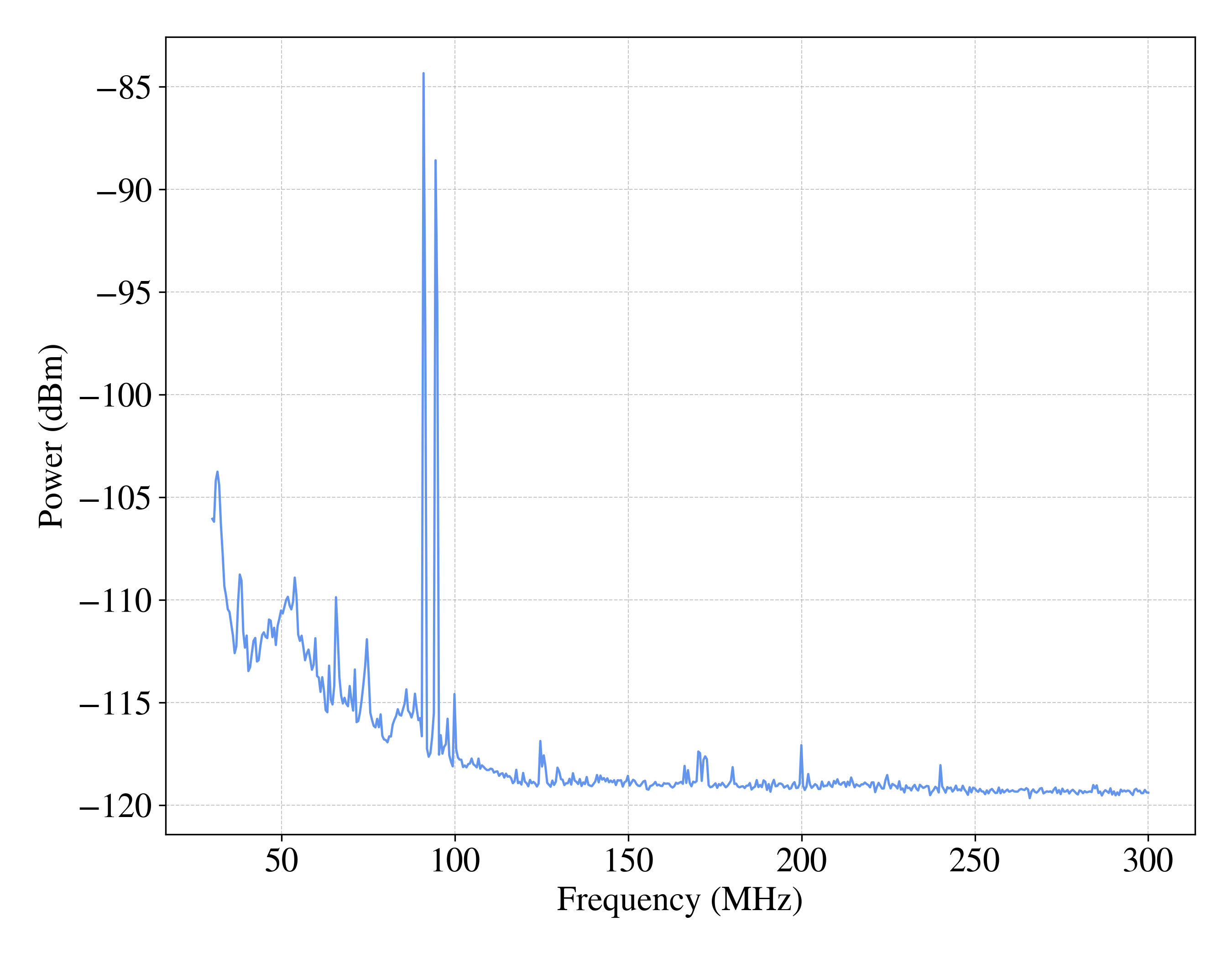}
        \caption{}
        \label{subfig:arctic_median}
    \end{subfigure}
    \caption{GLS drift scan highlighting RFI characteristics in the 30–300~MHz band. (a) Calibrated drift scan and (b) median along the temporal axis.}
    \label{fig:Arctic calibrated waterfalls}
\end{figure}

To understand the RFI signatures at each site, drift scans were generated from bandpass-calibrated data. These scans provide a clear and intuitive representation of the RFI environment. For example, transient RFI appears as short-duration spikes, whereas persistent RFI manifests as continuous vertical lines across the plot. Figures \ref{subfig:GBD_waterfall}, \ref{subfig:TLL_day_waterfall} and \ref{subfig:TLL_night_waterfall}, \ref{subfig:kalpong_waterfall}, and \ref{subfig:arctic_waterfall} present the drift scans for GRO, TLL (day and night), KDA and GLS, respectively. A preliminary visual inspection reveals the presence of both persistent and transient RFI at all sites. Additionally, broad-band features that vary between scans at different times were observed. 
To gain deeper insights into the RFI characteristics, we compute the median along the temporal axis. The resulting median plots are shown in Figures \ref{subfig:GBD_median}, \ref{subfig:TLL_day_median} and \ref{subfig:TLL_night_median}, \ref{subfig:kalpong_median}, and \ref{subfig:arctic_median}. These plots, presented in dBm units, provide a clear comparison of RFI presence and intensity across sites and frequency channels.

The visual analysis reveals both narrow-band spikes and broad-band features. Narrow-band spikes are typically associated with anthropogenic RFI signals. In contrast, the broad-band features require further investigation, as they may result from either RFI, the inherent shape of the sky spectrum, or uncalibrated system response.

\vspace{15pt}

\vspace{15pt}

\vspace{15pt}

\section{Algorithm for RFI detection} \label{sec:Data_analysis}

Obtaining an overview of the radio environment requires effective RFI detection techniques. Manual inspection of large datasets is impractical, particularly for surveys spanning multiple locations. To address this, two approaches were explored. The first approach involved an automated RFI detection algorithm implementing a Hampel filter, which is well-suited for identifying outliers. The second method involved employing singular value decomposition (SVD) of the drift scans, which works well for recognizing patterns in the data. More details about the algorithms involved will be elaborated in the later part of the section.

\subsection{RFI detection using Hampel Filter} \label{subsec:hample_approch}
\begin{figure}[h!]
    \centering
    \includegraphics[width=\textwidth]{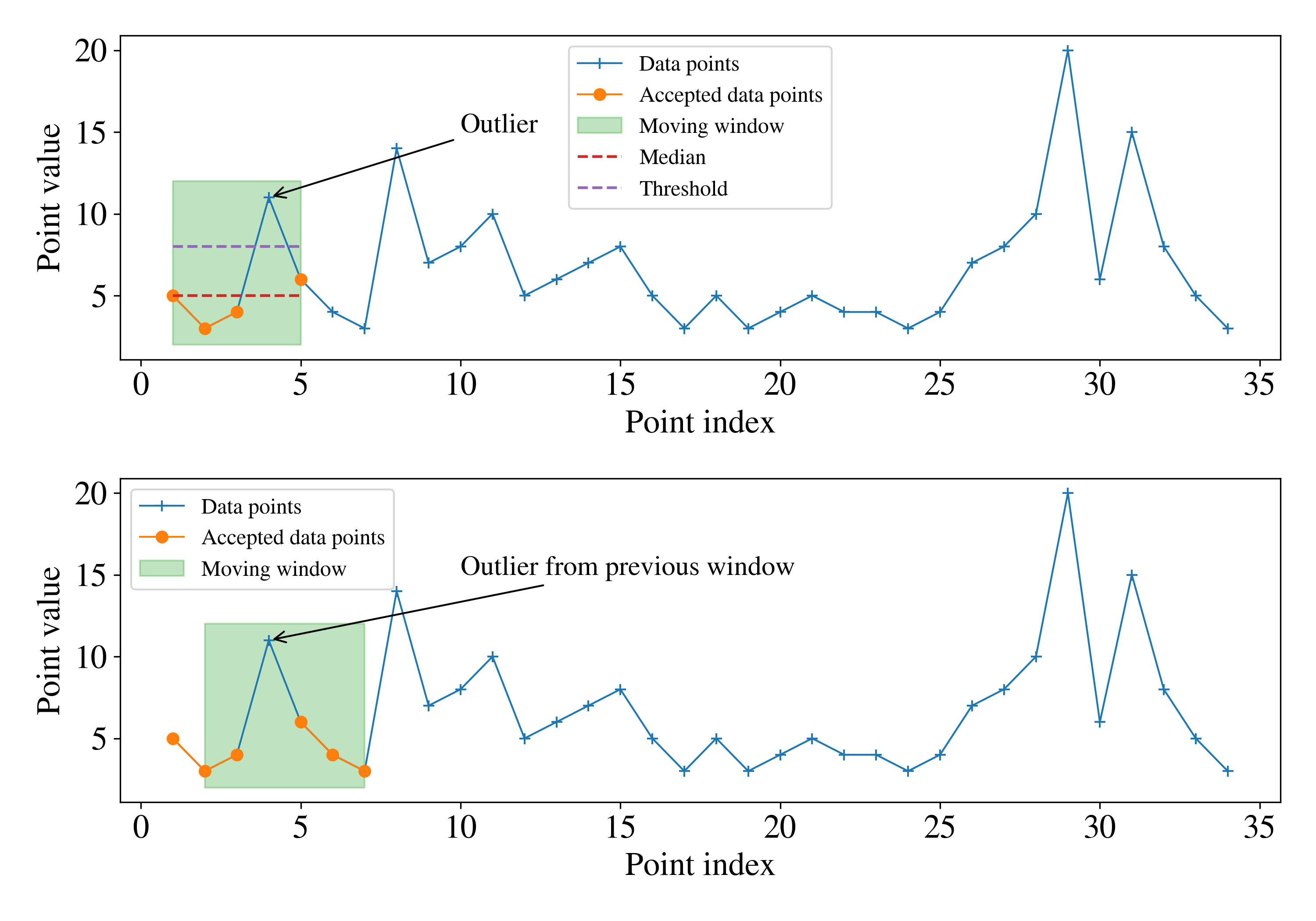}
    \caption{Demonstration of the working of a Hampel filter. The top panel shows the simulated data with a sliding window applied to detect outliers in that particular window. The bottom panel highlights the identified outliers, which are subsequently flagged. The window shifts by one point at each step and adjusts its size to include the same number of points, excluding any point previously flagged as an outlier.}
    \label{fig:Hampel_filter}
\end{figure}

The algorithm used here relies on the Hampel filter to detect and flag outliers\cite{DBLP:journals/corr/Suomela14}. It applies a sliding window operation with an adjustable width, as shown in Figure \ref{fig:Hampel_filter}. For each window, the median and median absolute deviation (MAD) of the points are calculated. Instead of directly calculating standard deviation from data, it utilizes the MAD to assess variability\cite{Rousseeuw01121993}. Outliers are identified as points that exceed a set threshold. The sliding window moves one element at a time, skipping previously flagged outliers. The window width is adaptive in that if it encounters flagged points, it adjusts its size to accommodate the same number of unflagged points in each iteration. The Hampel filter outperforms conventional filters like mean or Gaussian filters by using the median\cite{974338}, which makes it more robust against large deviations in data.  Additionally, the Hampel filter is both simple and computationally efficient, making it a robust tool for detecting outliers\cite{pearson2016generalized}.

The choice of window size and threshold in outlier detection algorithms like the Hampel filter significantly impacts accuracy. A smaller window size increases the filter’s sensitivity to noise by making threshold calculations less stable, which can result in falsely identifying small fluctuations as outliers and a high rate of false detections, especially in noisy datasets. Conversely, a larger window size stabilizes these calculations, smoothing over fine details but potentially missing low amplitude, weaker outliers by the biasing of the statistics due to stronger outliers in the same window. Similarly, the threshold level plays a crucial role. A lower threshold raises sensitivity, detecting minor deviations but risking many false detections by flagging natural variations as outliers. A higher threshold, on the other hand, decreases sensitivity, capturing only strong deviations and possibly missing small but meaningful outliers, such as faint but persistent signals, especially if they are just below the threshold. Balancing these parameters is essential to accurately detect RFI without excessive false detections or missed detections.

In the current analysis, a window size of 10 points is chosen, spanning approximately 5~MHz (10 points), where the threshold is taken to be 3 times the standard deviation (SD). This limited window size reduces the likelihood of overlapping multiple RFI sources. In frequency bands with closely packed and consistently occupied channels, such as the FM range\footnote{\href{https://www.itu.int/dms_pub/itu-r/opb/reg/R-REG-RR-2020-ZPF-E.zip}{https://www.itu.int}} (87.5-108~MHz) or the UHF-Satcom Satellites\footnote{\href{https://uhf-satcom.com/satellite-reception/uhf}{https://uhf-satcom.com/satellite-reception/uhf}} communication range (240-270~MHz), all data points are flagged prior to applying the Hampel filter. This preemptive flagging prevents these fully occupied bands from biasing the threshold. The Hampel filter is applied to each spectrum from the drift scan. For each frequency channel, the RFI occupancy is calculated as the fraction of spectra in which RFI in that channel is detected. This kind of filter is also prone to generating some false detections. In order to calculate the false detections fraction, drift scans of the noise floor were taken by connecting a 50-ohm termination load to the instrument, and the Hampel filter was run over the scans. The false detection fraction calculated as such is used to correct for the occupancy using

\begin{equation}
O_{\text{true}} = \frac{O_{\text{meas}} - f_{\text{false}}}{1 - f_{\text{false}}},
\label{eq: true occupancy}
\end{equation}

where $O_{\text{true}}$ and $O_{\text{meas}}$ are the true and measured occupancies respectively and $f_{\text{false}}$ is the false detection fraction.

RFI identified through this method is subsequently referred to as Hampel-filter-detected RFI. The occupancy for each site is shown in Figure \ref{fig: Occu_plots}. We present the results from this method in the subsequent section.

\subsection{RFI study using Singular Value decomposition}
The Hampel filter is a simple and efficient tool for detecting RFI, particularly narrow-band interference. However, its primary limitation lies in its inability to detect the presence of broad-band RFI in the spectrum, which remains effectively invisible to this method\cite{2001A&A...378..327F}. The reason is that broad-band RFI in a scan can blend with system shapes, which are broad-band structures in spectra and not appear as an outlier in the sample filter window. Singular Value Decomposition (SVD) excels at identifying patterns and separating distinct components in the data\cite{2018ApJ...853..187T}. SVD involves factorizing a given matrix A into three distinct matrices, such that,

\begin{equation}
    A = U \Sigma V^T,
    \label{eq:SVD}
\end{equation}

where $U$ and $V$ are orthogonal matrices containing the left and right singular vectors, respectively, and $\Sigma$ is a diagonal matrix of singular values\cite{1965SJNA....2..205G}.

This decomposition effectively identifies the intrinsic structure of the data by expressing it in terms of orthonormal basis vectors. The singular vectors represent different modes of variation in the dataset. These modes are ranked according to their corresponding singular values, which serve as an indicator of the amount of variance captured by each mode\cite{inbook}. Broad-band RFI in those scans, which may not be visible as an outlier to the Hampel filter, can still manifest as a prominent component in the singular values in SVD. While the first principal component captures the most persistent features, such as persistent narrow-band, broad-band RFI, and system shapes, consecutive prominent components efficiently capture the variability in the data. While persistent broad-band RFI can blend with the system shape, if it has temporal variability, it can be captured in subsequent components. An additional advantage offered by higher modes is the absence of any baseline spectrum shape contributed by the sky and any unmodeled bandpass. Therefore, RFI detection is more robust with a flat spectrum component.

\begin{figure}[h!]
    \centering
    \includegraphics[width=\textwidth]{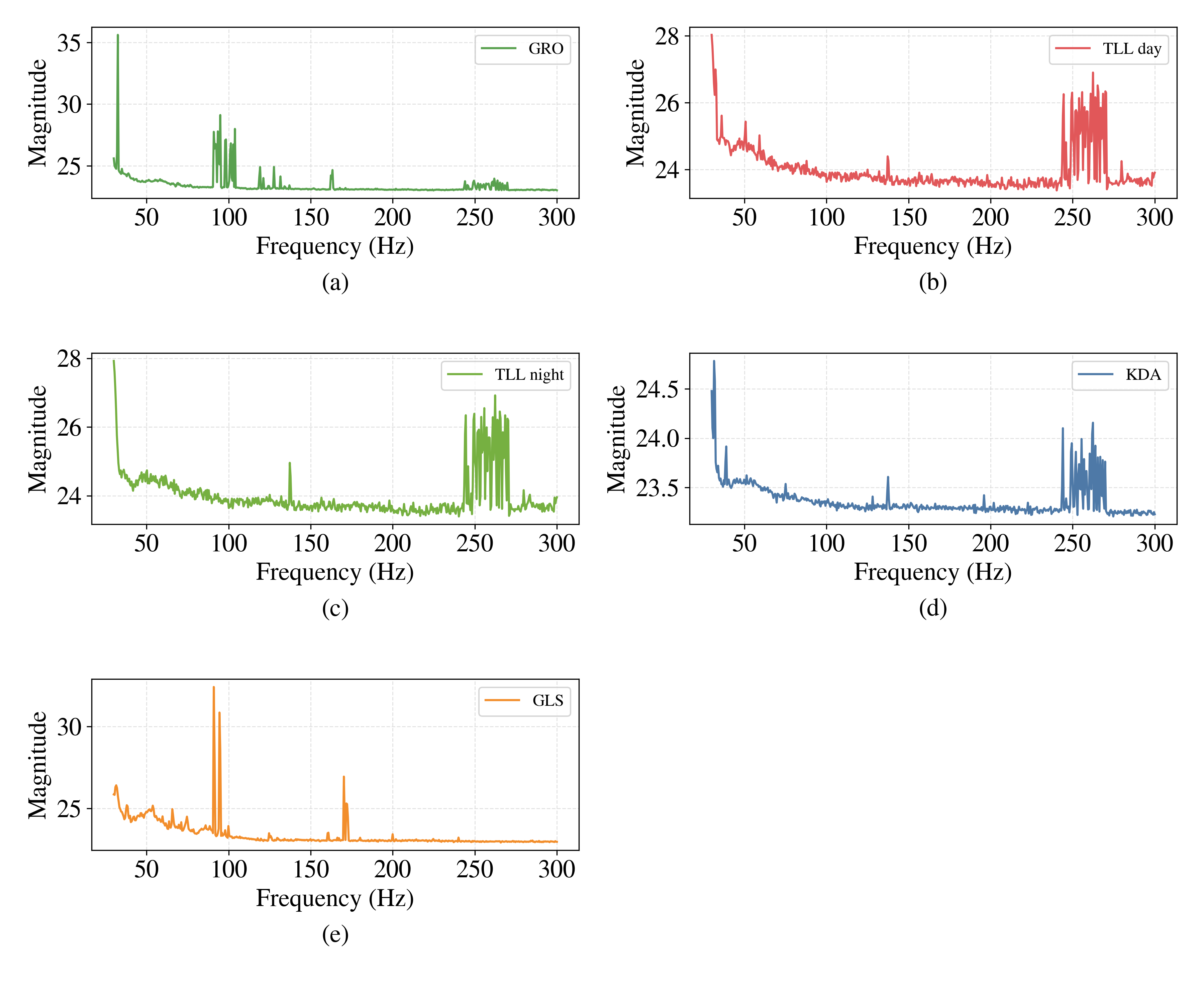}
    \caption{Principal Components for each site. (a) GRO, (b) TLL day, (c) TLL night, (d) KDA, (e) GLS.}
    \label{fig: principal components}
\end{figure}

For the analysis, the SVD of the drift scans was taken, and singular values were observed. The values of these scans are shown in Figure \ref{fig: singular values}. We also present a zoomed-in region that shows significant eigenvalues after the principal component before they saturate. The principal component captures the most prominent feature in the data. The principal component of each site is shown in Figure \ref{fig: principal components}. When compared to the median plots Figure \ref{fig:GBD calibrated waterfalls} (b),\ref{fig:Ladakh calibrated waterfalls} (b), \ref{fig:Kalpong calibrated waterfalls} (b) and \ref{fig:Arctic calibrated waterfalls} (b), the principal component clearly represent the most prominent features of the data. However, an analysis of consecutive modes will bring out additional significant features with increased variability.

Notably, the number of significant singular values varies across different sites, with the associated modes capturing substantial variability in the data reflected in these eigenvalues. To calculate the number of significant modes beyond the principal component, we compute the signal-to-noise ratio (SNR) of the detected outlier in each mode and note the number of modes at which the SNR stabilizes. Any further modes only capture amplitude variability in the already detected outlier frequency. The number of significant modes for each scan is given in Table \ref{tab: SVD details}. A similar Hampel filter, as described earlier in the Subsection \ref{subsec:hample_approch}, was applied to the principal component, and outliers were flagged as RFI. Since the baseline shape is represented in the principal mode, higher-order modes typically manifest as flat baselines, with RFI appearing as outliers. These outliers were also flagged using a median filter.

The flagged RFIs, identified through both the principal and significant higher-order modes, are collectively referred to as SVD-detected RFI. Since RFI detection is carried out in different modes, it is difficult to map their temporal occupancy with the mode in which it was detected. However, as mentioned earlier, this method is highly effective at identifying patterns, and the clustering of detected channels can reveal significant insights. Although broad-band RFI is often overlooked by the Hampel filter method, the clustering of channels identified through SVD can shed some light on it. Considering the window size of the Hampel filter was set to 10 points ($\sim$5~MHz), clusters are identified if more than five consecutive channels ($\sim$2.5~MHz) are flagged. Any group of five or more adjacent channels that are detected as RFI using the SVD method is marked as a cluster, and defined as a broad-band RFI. The spectral nature of `broad-band RFI' is discussed in detail in the Section \ref{sec: Results and Morphology of RFI}. We show broad-band RFI captured by SVD in Figure \ref{fig: BB_RFI}, with relevant details provided in Table \ref{tab: SVD details}.

\subsection{Comparison of RFI Detected Using Hampel Filter and Singular Value Decomposition}
The channels where RFI was detected using both methods were compared, revealing that while some channels were commonly identified, each method also had unique detections. This comparison is visually represented in Figure \ref{fig: SVD_Hampel_comparison}. Although the Hampel filter method provides a few exclusive detections, SVD identifies a significantly higher number of unique channels that have RFI. While the Hampel filter method captures the occupancy of narrow-band RFI, SVD offers valuable insights into the spectral characteristics of RFI in the drift scans. Furthermore, the number of significant modes highlights the complexity of the RFI environment. A more complex RFI environment implies a greater challenge in modeling the data for scientific analysis. Therefore, in the following section, in addition to discussing RFI morphology and occupancy, we also discuss overall site complexity indicated by the number of significant modes.

\begin{table}[h!]
\centering
\small 
\caption{Details of significant modes, Broad-band RFI (magnitude and clustered), and RFI detected exclusively by SVD.}
\label{tab: SVD details}
\begin{tabular}
{|p{0.1\textwidth}|p{0.15\textwidth}|p{0.15\textwidth}|p{0.15\textwidth}|p{0.25\textwidth}|}
\hline
\textbf{Site} & \textbf{Number of significant modes} & \multicolumn{2}{c|}{\textbf{broad-band RFI}} & \textbf{Channels captured exclusively by SVD (\%)} \\ \hline
 &  & \textbf{Clusters} & \textbf{percent of Bandwidth} &  \\ \hline
GRO & 60 & 10 & 40 & 23 \\ \hline
TLL day & 49 & 11 & 27 & 28 \\ \hline
TLL night & 31 & 7 & 19 & 21 \\ \hline
KDA & 22 & 4 & 23 & 12 \\ \hline
GLS & 53 & 5 & 26 & 31 \\ \hline
\end{tabular}

\end{table}

\section{Results} \label{sec: Results and Morphology of RFI}

Based on the survey conducted at these locations and the analysis of their drift scans, we find several known and unknown RFI-occupied channels. The most prominently  and identified occupied channels detected using a Hampel filter are listed in Table \ref{tab: occupancies}. The known sources mentioned are matched with the National Frequency Allocation document of the International Telecommunication Union\footnote{\href{https://www.itu.int/dms_pub/itu-r/opb/reg/R-REG-RR-2020-ZPF-E.zip}{https://www.itu.int}}. Apart from that, based on RFI detected using both Hampel filter and SVD, we classify RFI by both its occurrence and spectral characteristics.

\textbf{Occurrence}

The RFI observed in these scans can be classified as either persistent or transient based on its occurrence. Persistent RFI is present continuously throughout the scans, rendering these channels unusable for scientific analysis. By contrast, some RFI appears intermittently, either at periodic intervals or sporadically. Such RFI may originate from sources such as satellite downlink while it transits, meteor or aircraft reflections. In such cases, the occupancy of the RFI will depend on the number of passes that these sources make during the observation period.
The Hampel filter gives a robust estimate of the RFI occupancy.

\textbf{Spectral Characteristics}

RFI can be classified as either narrow-band or broad-band based on its spectral characteristics. Narrow-band RFI is more prevalent, as most transmission devices are confined to specific frequency channels. However, broad-band RFI can also occur, potentially due to transmitter malfunctions such as unintended emission from satellites\cite{2023A&A...676A..75D}\cite{2023A&A...678L...6G}, or an apparent shape observed due to radiometer saturation effects, which can render the entire band unusable\cite{2001A&A...378..327F}. High-density frequency allocations within a specific band, such as the FM radio band, can render the entire band unusable for scientific analysis. Although each transmission is individually narrow-band, their close spectral spacing causes them to appear as a broader feature in the data. For the purposes of this study, such features will also be referred to as broad-band RFI.

\textbf{Gauribidanur Radio Observatory, Karnataka, India}

\begin{figure}[h!]
    \centering
    \includegraphics[width=\textwidth]{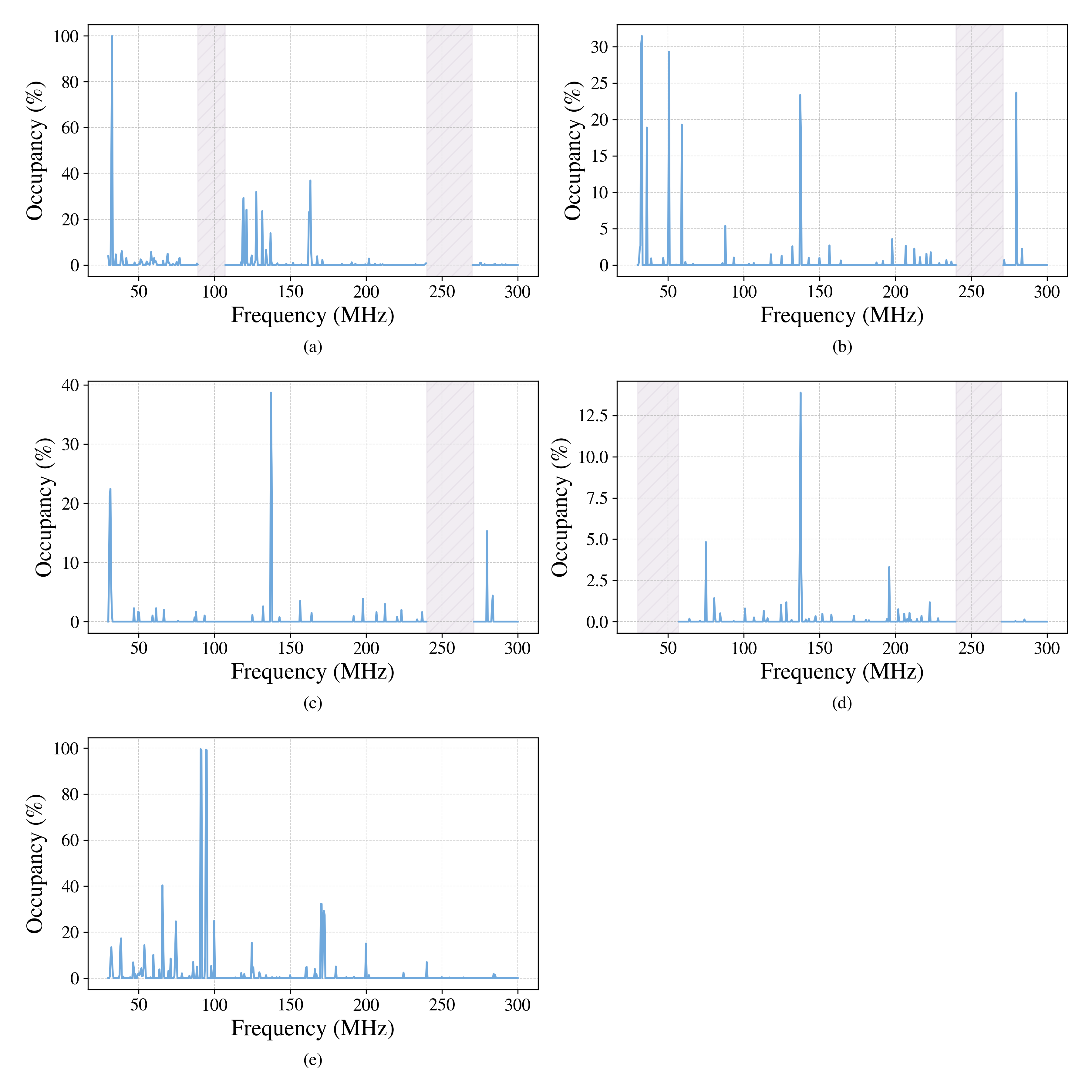}
    \caption{Occupancies of RFI present for each site in the 30–300~MHz frequency range. (a) GRO, (b) TLL day, (c) TLL night, (d) KDA, (e) GLS. The greyed-out regions are bands which are completely flagged for having 100 percent occupancy.}
    \label{fig: Occu_plots}
\end{figure}

\begin{figure}[h!]
    \centering
    \includegraphics[width=\textwidth]{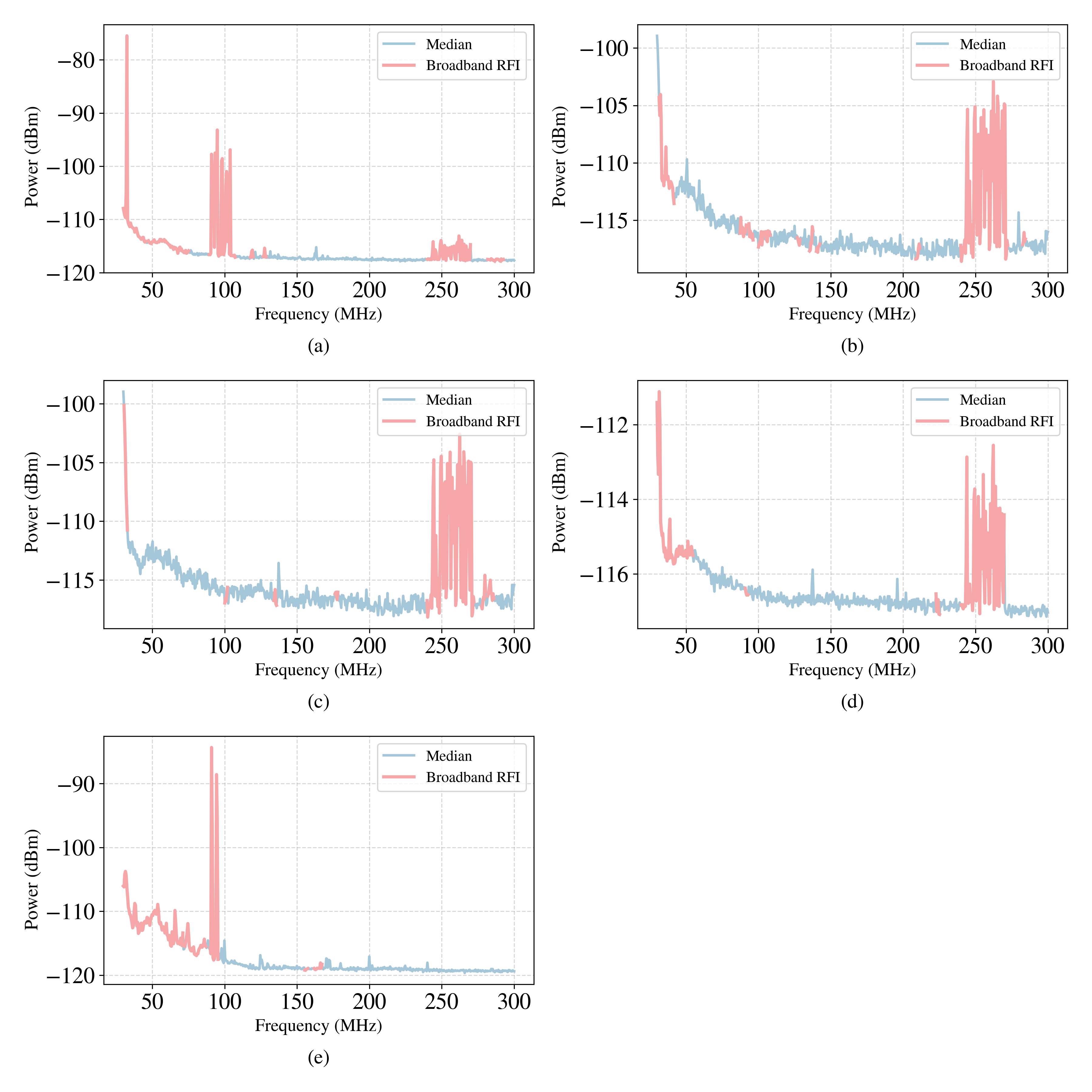}
    \caption{Broad-band RFI for each site. (a) GRO, (b) TLL day, (c) TLL night, (d) KDA, (e) GLS.}
    \label{fig: BB_RFI}
\end{figure}

\begin{figure}[h!]
    \centering
    \includegraphics[width=\textwidth]{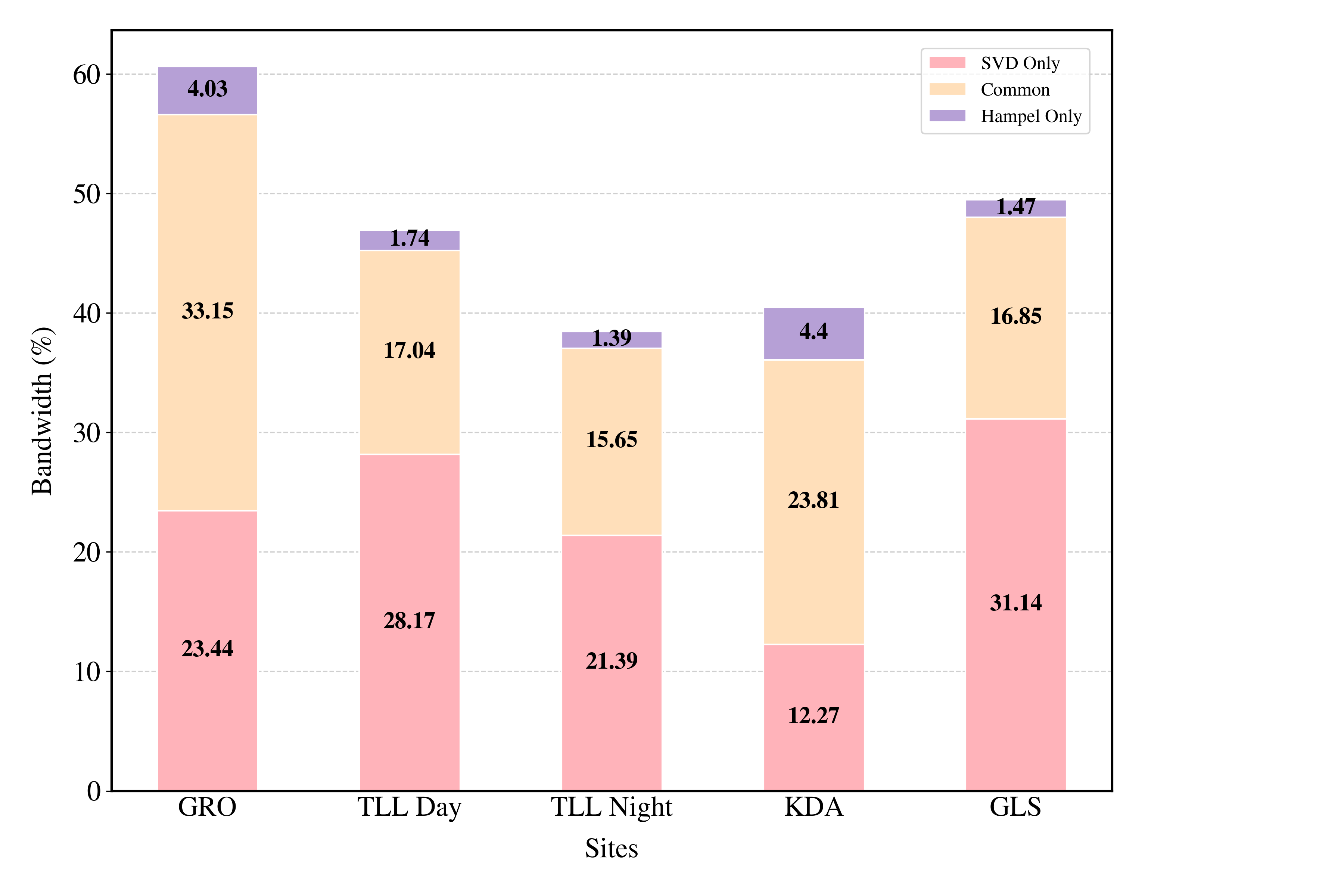}
    \caption{Comparison of the RFI detected exclusively by Hampel filter (Hampel Only) or SVD (SVD only), or the frequency channels that got detected by both methods (common) for having RFI. (left to right) GRO, TLL day, TLL night, KDA, GLS.}
    \label{fig: SVD_Hampel_comparison}
\end{figure}

The scans reveal multiple frequencies with varying occupancy levels. The bands at 87.5-108~MHz and 240-270~MHz were fully flagged, known to be allocated for FM radio and UHF-Satcom Satellites communications as clearly shown in Figure \ref{fig: Occu_plots} (a). Additionally, the 137.5~MHz downlink (DL) satellites show an occupancy of about $\sim$ 14~\%. As shown in Figure \ref{fig: Occu_plots} (a), we find several narrow-band RFI sources detected by algorithms described above. However, an SVD analysis provides a more detailed picture, revealing 60 significant modes indicating high variability in the strength and occupancy of RFI. As shown in Figure \ref{fig: SVD_Hampel_comparison}, approximately 61~\% of channels were flagged as RFI, with about 24~\% exclusively detected by the SVD method. Figure \ref{fig: BB_RFI} illustrates that the entire low-frequency band, from 30–75~MHz, is dominated by broad-band RFI, along with several additional patches of broad-band interference. Table \ref{tab: SVD details} summarizes the findings from the SVD analysis, which identified a total of 10 clusters of broad-band RFI, covering approximately 40~\% of the total band. The overall RFI coverage is the highest among all sites, with about 61~\% band covered with RFI at some point in time.

\textbf{Twin Lakes, Ladakh, India}

The TLL scans showed distinct patterns in the drift scans that correspond with the occupancy estimates. During the day, a low-frequency signal below 100~MHz appeared intermittently, likely due to reflections from passing aircraft\cite{Wilensky2019}. These signals were absent at night, aligning with reduced flight activity. Unlike GRO, no persistent RFI was detected in the FM band, but the 240-270~MHz band was continuously occupied by UHF-Satcom Satellites. 137.5~MHz DL satellites were also active , with occupancy below 40~\%. The night scans otherwise showed clean bands from 40-240~MHz with most RFI occupancy less than 5~\%, with 137.5~MHz DL satellites as the only notable RFI source. As per the SVD analysis, given in Table \ref{tab: SVD details}, 49 significant modes were found for the day and only 31 for the night, indicating a more stable RFI environment compared to GRO. TLL night has 7 clusters of broad-band RFI and a broad-band cover of 19~\%, the least among all sites. TLL day, however, has 11 clusters and an increased broad-band cover of 27~\%. RFI levels are relatively low, leaving more than 60~\% of the band available for scientific observation, considering only channels with no RFI detections at any time.

\textbf{Kalpong Dam Andaman, India}

\begin{figure}[h!]
    \centering
    \includegraphics[width=\textwidth]{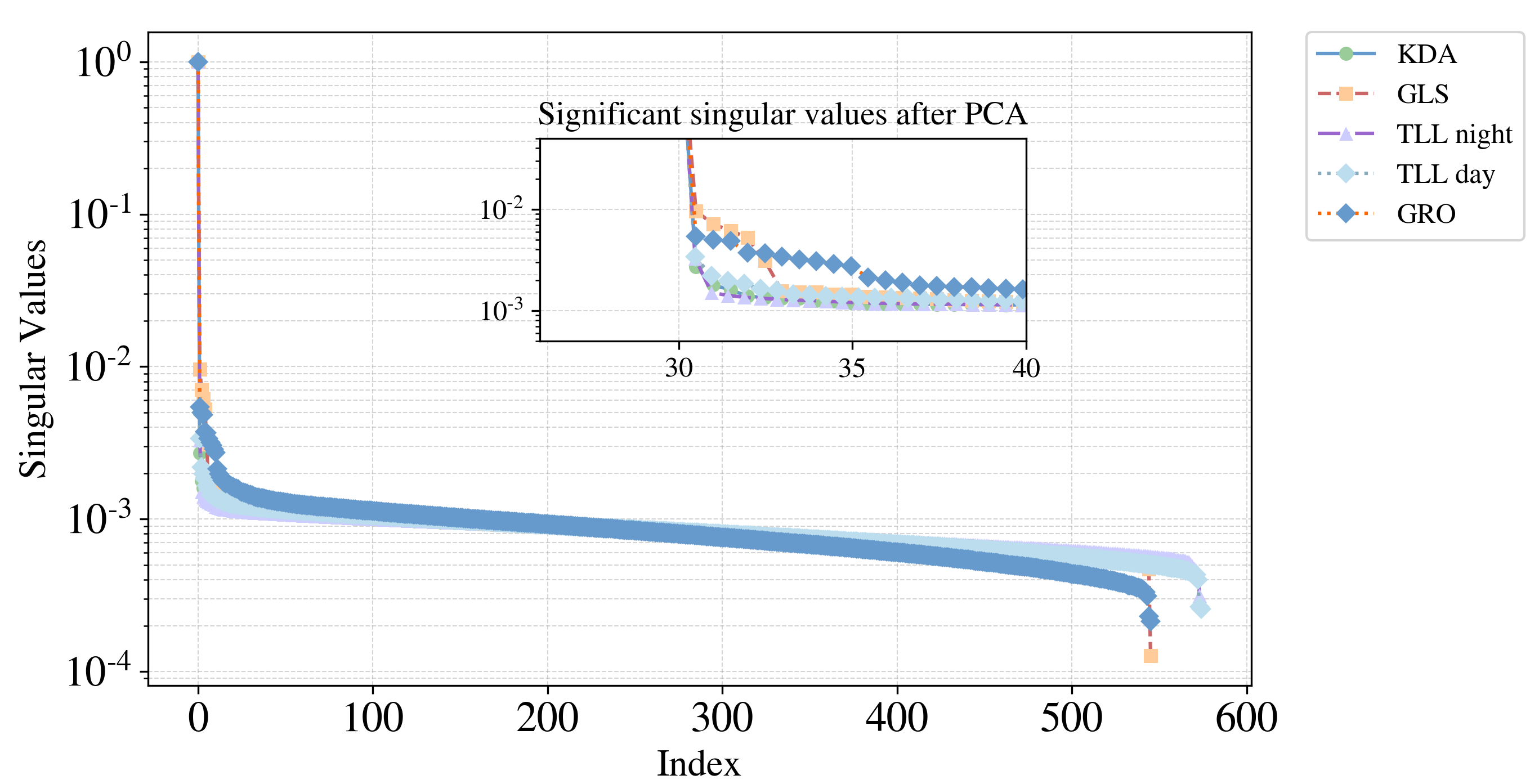} 
    \caption{Singular values (normalized) for each site. (a) GRO, (b) TLL day, (c) TLL night, (d) KDA, (e) GLS. The zoomed-in plot captures the significant values post the value of the principal component.}
    \label{fig: singular values}
\end{figure}

As observed in the drift scans, the frequency band below 55~MHz is strongly affected by RFI and was fully flagged prior to applying the Hampel filter. The 240-270~MHz range shows consistent occupancy by UHF-Satcom Satellites, and 137.5~MHz DL satellites appear again with occupancy $\sim$13~\%. Other than these sources, the band from 55-240~MHz appears largely clean and free from RFI, with most narrow-band RFI occupancy below $\sim$5~\%. Post-SVD analysis, we find that it has the least number of significant modes, only 22, which is also evident in Figure \ref{fig: singular values} where KDA singular values stabilize quicker than the rest of the sites. Figure \ref{fig: SVD_Hampel_comparison} shows that KDA has the second least occupied channels with RFI, with only $\sim$12~\% RFI channels exclusively detected by SVD analysis. Further analysis shows that it also has 4 clusters, making 22~\% bandwidth covered with broad-band RFI. The overall RFI of around 40~\%.

\textbf{Gruvebadet Atmosphere Laboratory, Ny-Ålesund, Svalbard, Norway}

The band below 100~MHz seems to be dominated by RFI, as evident from the drift scan plot, with two distinct FM radio channels, 91 and 94~MHz, with 100~\% occupancy. Frequency bands higher than 100~MHz appear to be clean based on occupancies calculated using the Hampel filter, where the only notable RFI is at 171-172~MHz with occupancies below 40~\%. This is a known frequency used for local wireless communication. Apart from that, we observe some low occupancy RFIs below 20~\%. It should be noted that 137.5~MHz DL satellites are not detected at this location. SVD analysis shows that it has 53 significant modes, second only to GRO. With 5 clusters, it has a 26~\% of band detected with broad-band RFI. This site also has the highest percentage of channels that are exclusively detected by the SVD method, which is $\sim$31~\%. The overall RFI coverage for this site is $\sim$49~\% but mostly limited to frequencies below 100~MHz.

\begin{table}[htbp]
\centering
\caption{Notable RFI sources identified across different frequency channels at the surveyed sites, matched with allocations from the International Telecommunication Union’s national frequency allocation table\tablefootnote{\href{https://www.itu.int/dms_pub/itu-r/opb/reg/R-REG-RR-2020-ZPF-E.zip}{https://www.itu.int}}.}
\label{tab: occupancies}
\begin{tabular}{|>{\centering\arraybackslash}p{3cm}|
                >{\centering\arraybackslash}p{4.5cm}|
                >{\centering\arraybackslash}p{3cm}|}
\hline
\textbf{Frequency (MHz)} & \textbf{Source} & \textbf{Occupancy (\%)} \\ \hline

\rowcolor{gray!20} \multicolumn{3}{|c|}{\textbf{Gauribidanur Radio Observatory, Karnataka}} \\ \hline
87.5--108 & FM Radio & 100 \\ \hline
137.5 & 137.5 MHz Downlink satellites & 14 \\ \hline
240--270 & UHF-Satcom Satellites & 100 \\ \hline

\rowcolor{gray!20} \multicolumn{3}{|c|}{\textbf{Ladakh Twin Lakes (Day)}} \\ \hline
137.5 & 137.5 MHz DL satellites & 23 \\ \hline
240--270 & UHF-Satcom Satellites & 100 \\ \hline

\rowcolor{gray!20} \multicolumn{3}{|c|}{\textbf{Ladakh Twin Lakes (Night)}} \\ \hline
137.5 & 137.5 MHz DL satellites & 38 \\ \hline
240--270 & UHF-Satcom Satellites & 100 \\ \hline

\rowcolor{gray!20} \multicolumn{3}{|c|}{\textbf{Kalpong Dam (Andaman)}} \\ \hline
137.5 & Satellites with 137.5 MHz DL & 14 \\ \hline
240--270 & UHF-Satcom Satellites & 100 \\ \hline

\rowcolor{gray!20} \multicolumn{3}{|c|}{\textbf{Gruvebadet Atmosphere Lab, Ny-Ålesund, Norway}} \\ \hline
91 and 95 & FM Radio & 100 \\ \hline
172 & Local FM communication & 29 \\ \hline

\end{tabular}
\end{table}

\section{Discussions and Conclusions} \label{sec:Discussion and Conclusions}
This work addresses diverse aspects of site surveys for the goal of radio experiment deployment. Although the study's applications are broad, this study is more centred around radio experiments related to cosmology, such as a global 21-cm signal, particularly with a SARAS-like radiometer. Selecting an observation site requires a thorough understanding of multiple site parameters and their impact on data quality. In this study, we consider key factors that influence the deployment of instruments for radio observations. The parameters that are reported here are the horizon profile, the observable sky, and most importantly, the frequency bands available for observations.  

The instrument for the RFI survey was designed for its portability and ease of deployment.
 The entire system can be assembled within minutes. Its performance remains uniform across sites, as confirmed by reflection coefficient measurements, and the simulations align well with the site measurements.

Several frequency structures are evident across all drift scans and surveyed sites, becoming particularly prominent upon visual inspection of the median plots (Figures \ref{fig:GBD calibrated waterfalls} (b), \ref{fig:Ladakh calibrated waterfalls} (b) and (d), \ref{fig:Kalpong calibrated waterfalls} (b), and \ref{fig:Arctic calibrated waterfalls} (b)). Understanding their origin and their spectro-temporal nature would be crucial for any scientific deployment. This was achieved through occupancy analysis using a Hampel filter that shed light on the narrow-band RFI environment of each of the drift scans at the surveyed sites. The RFI occupancy analysis quantifies interference across scans, offering a measure of the extent to which each frequency channel is affected. Since the resolution of each frequency channel is $\sim$500~KHz, a very narrow-band RFI will suffer band smearing. While the scans are recorded at a time resolution of approximately 2 seconds, any short-duration RFI signal which appears in between two scans has the possibility of being missed. However, their transient nature means they contribute negligibly to the RFI occupancy measurement of the channel. 
Each site shows unique occupancy profiles as shown in Figure \ref{fig: Occu_plots}. To further our understanding of the drift scans, the SVD method was employed. As mentioned, SVD analysis is useful in identifying patterns and variabilities in the data; we see that it detected several features that were missed by the Hampel filter method. While the outliers identified in significant SVD modes do not directly indicate occupancies, their presence in the most significant modes suggests they are prominent within the scan. These features can be a major influence on the quality of the data. The SVD method exploited the possible variability in the broad-band RFI to distinguish those features from the systematic and inherent foregrounds. They detected broad-band features remarkably well in the scans. A complete understanding of a site's suitability requires using both the Hampel filter and SVD methods. Based on these analyses and their results, the following conclusions can be drawn for each site.

\begin{itemize}
    \item \textbf{GRO}: 
    We see that amongst all sites, this site has the least obstruction on the horizon, and its latitude of $13.6^\circ$ N allowed more than 90~\% of the sky to be observable. Its horizon will offer the least complexity in modeling. However, this also made it very susceptible to terrestrial RFI, as evident from the occupancies in the Figure \ref{fig: Occu_plots} (a) and broad-band RFI Figure \ref{fig: BB_RFI} (a), especially due to its proximity to urban regions. Frequencies from 87.5-108~MHz and 240-270~MHz, used for FM radio and UHF-Satcom Satellites communication, were fully flagged due to 100 percent occupancy. The 137.5~MHz DL satellites showed moderate occupancy, around $\sim$14~\%. Given the loud radio environment, reaching the sensitivity required for CD/EOR 21~cm experiments might be challenging.
    
    \item \textbf{TLL}: 
    The site location had a horizon profile that obstructs around $0.0068\pi$ str sky with the highest peak of about $11^\circ$. The latitude of this location, being further up north at $32.97^\circ$ N, makes around 20~\% of sky always below horizon as seen in Figure \ref{fig: foregrounds}. Notable daytime RFI below 100~MHz likely resulted from airplane reflections and ongoing activities such as road construction in the region, while night scans were free of this interference, aligning with reduced flight activity in the region. FM band RFI was very rare here, but the 240-270~MHz range remained fully occupied by UHF-Satcom Satellites. 137.5~MHz DL satellites occupancy here at night was below 40~\%. Based on observed occupancy levels, nighttime at this location offers the quietest RFI environment among all sites. Besides 137.5~MHz DL satellites and a few clusters of broad-band RFI, which make up less than 20~\% of the entire band, the site remains radio-quiet.
    This makes it an ideal site for radio observation in the entire band of 40-200~MHz.
    
    \item \textbf{KDA}:
    The site location had a horizon profile that obstructs around $0.0126\pi$ str sky with the highest peak of about $12.5^\circ$. The latitude of this location lies close to that of GRO at $13.1^\circ$ N, enabling around 90~\% of the sky for observation.
    Frequencies below 55~MHz were entirely flagged, while the 240-270~MHz range showed consistent occupancy of 100~\%. UHF-Satcom Satellites had lower occupancy, around 13~\%, and the majority of the 55-240~MHz range appeared RFI-free except a few lines having occupancies below 5~\%. This site also offers a great opportunity for the deployment of instruments for radio astronomical observations in frequency ranges above 55~MHz. 
    
    \item \textbf{GLS}: 
     The site location had a horizon profile that obstructs most, around $0.0141\pi$ str sky, with the highest peak of about $17^\circ$. The latitude of this location is further up north at $78.9^\circ$ N,  keeping most of the southern sky completely below the horizon as seen in Figure \ref{fig: foregrounds}. Strong RFI below 100~MHz was observed, presumably from other equipment operating in the vicinity of the radiometer. We see full occupancy for FM channels at 91 and 94~MHz. Frequency channels above 100~MHz were cleaner, with notable RFI only at 171-172~MHz, though it remains below 40~\%. We do not see any significant clusters of broad-band RFI beyond 100~MHz, making this site an ideal candidate for observations above 100~MHz.
    
\end{itemize}

\subsection{Impact of RFI on SARAS system performance} \label{sec:SARAS impact}
The SARAS experiment seeks to detect the global 21-cm signal from the cosmic dawn and the epoch of reionization, which is expected to lie within the 40-200~MHz frequency range. As we find in the survey, this band is shared by several other transmitters, including FM radio (87.5-108~MHz), 137.5~MHz DL satellites, etc., in addition to several out-of-band RFI. The presence of these transmissions varies by location and poses challenges for SARAS observations. Conducting such surveys is essential for developing observation strategies and data modeling methodologies for such experiments. The SARAS antenna also has a wide, lifted toroidal beam, similar to the portable antenna that we used for the survey. This feature does make it susceptible to terrestrial RFI and the horizon profile. 

At the four surveyed sites, we identified sections of the 40–200~MHz band that were suitable for scientific observations. The SARAS antenna, with its wide beam, will capture multiple RFI sources at GRO, making this site highly unsuitable for observations in its operation range of 40-200~MHz. TLL, on the other hand, provides the most favourable conditions, especially at night, with minimal interference across the band. KDA also presents a promising location; in spite of its horizon complexity, most of the 55–200~MHz range is interference-free. Here, SARAS can operate effectively above 55~MHz without significant contamination from terrestrial sources, making it a viable deployment site. GLS, despite cleaner conditions above 100~MHz, suffers from heavy RFI contamination below 100~MHz, limiting the observing frequency band. Given these considerations, SARAS is best suited for deployment at TLL and KDA, where the combination of sky accessibility and minimal RFI allows for effective wideband observations. GLS also provides a suitable site for deployment, though with a limited band.

 While the parameters considered here are specifically tailored for CD/EoR experiments, the methodology and algorithms can be broadly applied to a wide range of observational deployments.

\section*{Acknowledgements}
This work is funded by the core research grant under the Science and Engineering Research Board, Department of Science and Technology, India.

The authors extend their heartfelt gratitude to Santosh Harish for his efforts in porting the data acquisition program from Windows to Linux. We sincerely appreciate the contributions of Kasturi S. (Electronics Engineering Group, Raman Research Institute) and Mohamed Ibrahim and his team in the Mechanical Engineering Services (Raman Research Institute).

As the agency responsible for India’s research activities in the polar regions, we thank NCPOR, Goa, for the unique opportunity to characterize the radio frequency environment in and around Ny-Ålesund, Svalbard, Norway. Thanks to NCPOR for providing the resources and facilities necessary for this endeavour. Special thanks to Dr. Manish Tiwari, Dr. Rohit Srivastava, and Dr. Nuncio Murukesh for their unwavering support during various phases of this work.

The authors are especially grateful to Dr. Utpal Sharma, Special Secretary (IT), Andaman and Nicobar (A\&N) Administration, for his steadfast support, which was instrumental in the smooth execution of this endeavour at the Kalpong Dam. His continuous assistance greatly facilitated the success of this activity. We also owe gratitude to Shri B. Mohan Babu, State Project Manager, SOVTECH, A\&N Administration, for the exceptional technical and logistical support provided, even during odd hours and during holidays. Special thanks are due to the Superintending Engineer and other engineers of the Electricity department, A\&N Islands, for all the logistical support extended during our RFI monitoring campaign in the Kalpong Dam.

The contributions of Mr. Dorje Angchuk, Engineer-in-charge, IAO, Leh, and staff at the Indian Astronomical Observatory, Hanle, in providing logistics and technical support are gratefully acknowledged.
GBS deeply appreciates Ms. Julie Tromborg, Research Adviser, Kings Bay AS, for her unwavering support throughout the RFI monitoring campaign in Ny-Ålesund. Special thanks to Benedikt Uherek, Erlend Havenstrom, and Ida Kristoffersen for their invaluable assistance in swiftly setting up the RFI recording system on the Gruvebadet rooftop. GBS would like to sincerely thank Hege Iren Svåsand Kråkevik, Arnt Rennan, and Ingrid Kjerstad of the Norwegian Polar Institute for their crucial role in operations and logistics.

The authors would like to thank Ravi Subrahmanyan and Jishnu Nambissan Thekkeppattu for their fruitful comments and suggestions.

Finally, the authors would like to thank members of the RRI administration for providing the necessary resources and support for the smooth execution of our RFI monitoring campaigns. Finally, we acknowledge the contributions of the Gauribidanur staff, whose efforts were invaluable in the completion of this work.

\bibliography{bibfile}

@ARTICLE{2021ApJ...923...33B,
       author = {{Bassett}, Neil and {Rapetti}, David and {Tauscher}, Keith and {Nhan}, Bang D. and {Bordenave}, David D. and {Hibbard}, Joshua J. and {Burns}, Jack O.},
        title = "{Lost Horizon: Quantifying the Effect of Local Topography on Global 21 cm Cosmology Data Analysis}",
      journal = {\apj},
         year = 2021,
        month = dec,
       volume = {923},
       number = {1},
          eid = {33},
        pages = {33},
          doi = {10.3847/1538-4357/ac1cde},
archivePrefix = {arXiv},
       eprint = {2106.02153},
 primaryClass = {astro-ph.CO},
       adsurl = {https://ui.adsabs.harvard.edu/abs/2021ApJ...923...33B}
}

@article{t2021saras,
  author       = {Jishnu Nambissan T. and Ravi Subrahmanyan and R. Somashekar and N. Udaya Shankar and Saurabh Singh and A. Raghunathan and B. S. Girish and K. S. Srivani and Mayuri Sathyanarayana Rao},
  title        = {{SARAS 3 CD/EoR radiometer: design and performance of the receiver}},
  journal      = {Experimental Astronomy},
  volume       = {51},
  number       = {2},
  pages        = {193--234},
  year         = {2021},
  doi          = {10.1007/s10686-020-09697-2},
  url          = {https://doi.org/10.1007/s10686-020-09697-2},
  issn         = {1572-9508},
}

@article{1999A&A...345..380S,
	Adsurl = {http://adsabs.harvard.edu/abs/1999A%26A...345..380S},
	Author = {{Shaver}, P.~A. and {Windhorst}, R.~A. and {Madau}, P. and {de Bruyn}, A.~G.},
	Eprint = {astro-ph/9901320},
	Journal = {Astronomy and Astrophysics},
	Month = may,
	Pages = {380-390},
	Title = {{Can the reionization epoch be detected as a global signature in the cosmic background?}},
	Volume = 345,
	Year = 1999}

@article{2006PhR...433..181F,
	Adsurl = {http://adsabs.harvard.edu/abs/2006PhR...433..181F},
	Author = {{Furlanetto}, S.~R. and {Oh}, S.~P. and {Briggs}, F.~H.},
	Doi = {10.1016/j.physrep.2006.08.002},
	Eprint = {astro-ph/0608032},
	Journal = {\physrep},
	Month = oct,
	Pages = {181-301},
	Title = {{Cosmology at low frequencies: The 21 cm transition and the high-redshift Universe}},
	Volume = 433,
	Year = 2006}

@article{2012RPPh...75h6901P,
	Adsurl = {http://adsabs.harvard.edu/abs/2012RPPh...75h6901P},
	Archiveprefix = {arXiv},
	Author = {{Pritchard}, J.~R. and {Loeb}, A.},
	Doi = {10.1088/0034-4885/75/8/086901},
	Eid = {086901},
	Eprint = {1109.6012},
	Journal = {Reports on Progress in Physics},
	Month = aug,
	Number = 8,
	Pages = {086901},
	Title = {{21 cm cosmology in the 21st century}},
	Volume = 75,
	Year = 2012}

@ARTICLE{2004MNRAS.347..187F,
       author = {{Furlanetto}, Steven R. and {Sokasian}, Aaron and {Hernquist}, Lars},
        title = "{Observing the reionization epoch through 21-centimetre radiation}",
      journal = {\mnras},
         year = 2004,
        month = jan,
       volume = {347},
       number = {1},
        pages = {187-195},
          doi = {10.1111/j.1365-2966.2004.07187.x},
archivePrefix = {arXiv},
       eprint = {astro-ph/0305065},
 primaryClass = {astro-ph},
       adsurl = {https://ui.adsabs.harvard.edu/abs/2004MNRAS.347..187F}
}

@article{Rousseeuw01121993,
author = {Peter J. Rousseeuw and Christophe Croux},
title = {Alternatives to the Median Absolute Deviation},
journal = {Journal of the American Statistical Association},
volume = {88},
number = {424},
pages = {1273--1283},
year = {1993},
publisher = {ASA Website},
doi = {10.1080/01621459.1993.10476408},
}

@ARTICLE{2024MNRAS.527.2413P,
       author = {{Pattison}, Joe H.~N. and {Anstey}, Dominic J. and {de Lera Acedo}, Eloy},
        title = "{Modelling a hot horizon in global 21-cm experimental foregrounds}",
      journal = {\mnras},
         year = 2024,
        month = jan,
       volume = {527},
       number = {2},
        pages = {2413-2425},
          doi = {10.1093/mnras/stad3378},
archivePrefix = {arXiv},
       eprint = {2307.02908},
 primaryClass = {astro-ph.CO},
       adsurl = {https://ui.adsabs.harvard.edu/abs/2024MNRAS.527.2413P}
}

@ARTICLE{2008MNRAS.388..247D,
       author = {{de Oliveira-Costa}, Ang{\'e}lica and {Tegmark}, Max and {Gaensler}, B.~M. and {Jonas}, Justin and {Landecker}, T.~L. and {Reich}, Patricia},
        title = "{A model of diffuse Galactic radio emission from 10 MHz to 100 GHz}",
      journal = {\mnras},
         year = 2008,
        month = jul,
       volume = {388},
       number = {1},
        pages = {247-260},
          doi = {10.1111/j.1365-2966.2008.13376.x},
archivePrefix = {arXiv},
       eprint = {0802.1525},
 primaryClass = {astro-ph},
       adsurl = {https://ui.adsabs.harvard.edu/abs/2008MNRAS.388..247D}
}

@ARTICLE{974338,
  author={Pearson, R.K.},
  journal={IEEE Transactions on Control Systems Technology}, 
  title={Outliers in process modeling and identification}, 
  year={2002},
  volume={10},
  number={1},
  pages={55-63},
  doi={10.1109/87.974338}}

@article{DBLP:journals/corr/Suomela14,
  author       = {Jukka Suomela},
  title        = {Median Filtering is Equivalent to Sorting},
  journal      = {CoRR},
  volume       = {abs/1406.1717},
  year         = {2014},
  url          = {http://arxiv.org/abs/1406.1717},
  eprinttype    = {arXiv},
  eprint       = {1406.1717},
  bibsource    = {dblp computer science bibliography, https://dblp.org}
}

@inproceedings{inproceedings,
author = {Baan, Willem},
year = {2011},
month = {09},
pages = {1 - 2},
title = {RFI mitigation in radio astronomy},
doi = {10.1109/URSIGASS.2011.6051248}
}

@inbook{inbook,
author = {Bezina, Václav},
year = {2025},
month = {01},
pages = {75-93},
title = {Singular Value Decomposition},
isbn = {978-3-031-68645-0},
doi = {10.1007/978-3-031-68646-7_5}
}

@ARTICLE{2020MNRAS.498..265W,
       author = {{Wilensky}, Michael J. and {Barry}, Nichole and {Morales}, Miguel F. and {Hazelton}, Bryna J. and {Byrne}, Ruby},
        title = "{Quantifying excess power from radio frequency interference in Epoch of Reionization measurements}",
      journal = {\mnras},
         year = 2020,
        month = oct,
       volume = {498},
       number = {1},
        pages = {265-275},
          doi = {10.1093/mnras/staa2442},
archivePrefix = {arXiv},
       eprint = {2004.07819},
 primaryClass = {astro-ph.IM},
       adsurl = {https://ui.adsabs.harvard.edu/abs/2020MNRAS.498..265W}
}

@ARTICLE{2021ITAP...69.6209R,
       author = {{Raghunathan}, Agaram and {Subrahmanyan}, Ravi and {Shankar}, N. Udaya and {Singh}, Saurabh and {Nambissan}, Jishnu and {Kavitha}, K. and {Mahesh}, Nivedita and {Somashekar}, R. and {Sindhu}, Gaddam and {Girish}, B.~S. and {Srivani}, K.~S. and {Rao}, Mayuri S.},
        title = "{A Floating Octave Bandwidth Cone-Disk Antenna for Detection of Cosmic Dawn}",
      journal = {IEEE Transactions on Antennas and Propagation},
         year = 2021,
        month = oct,
       volume = {69},
       number = {10},
        pages = {6209-6217},
          doi = {10.1109/TAP.2021.3069563},
       adsurl = {https://ui.adsabs.harvard.edu/abs/2021ITAP...69.6209R}
}

@article{761c98bcdd13467f87dd40e7180307b3,
title = "Sky-averaged 21-cm signal extraction using multiple antennas with an SVD framework: the REACH case",
author = "Anchal Saxena and Meerburg, {P. Daniel} and Acedo, {Eloy de Lera} and Will Handley and Koopmans, {L{\'e}on V. E.}",
note = "11 pages, 13 figures. Published in MNRAS",
year = "2023",
month = jun,
day = "1",
doi = "10.1093/mnras/stad1047",
language = "English",
volume = "522",
pages = "1022–1032",
journal = "Monthly Notices of the Royal Astronomical Society",
issn = "0035-8711",
publisher = "Oxford University Press",
number = "1",
}

@ARTICLE{2013MNRAS.435..584O,
       author = {{Offringa}, A.~R. and {de Bruyn}, A.~G. and {Zaroubi}, S. and {Koopmans}, L.~V.~E. and {Wijnholds}, S.~J. and {Abdalla}, F.~B. and {Brouw}, W.~N. and {Ciardi}, B. and {Iliev}, I.~T. and {Harker}, G.~J.~A. and {Mellema}, G. and {Bernardi}, G. and {Zarka}, P. and {Ghosh}, A. and {Alexov}, A. and {Anderson}, J. and {Asgekar}, A. and {Avruch}, I.~M. and {Beck}, R. and {Bell}, M.~E. and {Bell}, M.~R. and {Bentum}, M.~J. and {Best}, P. and {B{\^\i}rzan}, L. and {Breitling}, F. and {Broderick}, J. and {Br{\"u}ggen}, M. and {Butcher}, H.~R. and {de Gasperin}, F. and {de Geus}, E. and {de Vos}, M. and {Duscha}, S. and {Eisl{\"o}ffel}, J. and {Fallows}, R.~A. and {Ferrari}, C. and {Frieswijk}, W. and {Garrett}, M.~A. and {Grie{\ss}meier}, J. and {Hassall}, T.~E. and {Horneffer}, A. and {Iacobelli}, M. and {Juette}, E. and {Karastergiou}, A. and {Klijn}, W. and {Kondratiev}, V.~I. and {Kuniyoshi}, M. and {Kuper}, G. and {van Leeuwen}, J. and {Loose}, M. and {Maat}, P. and {Macario}, G. and {Mann}, G. and {McKean}, J.~P. and {Meulman}, H. and {Norden}, M.~J. and {Orru}, E. and {Paas}, H. and {Pandey-Pommier}, M. and {Pizzo}, R. and {Polatidis}, A.~G. and {Rafferty}, D. and {Reich}, W. and {van Nieuwpoort}, R. and {R{\"o}ttgering}, H. and {Scaife}, A.~M.~M. and {Sluman}, J. and {Smirnov}, O. and {Sobey}, C. and {Tagger}, M. and {Tang}, Y. and {Tasse}, C. and {Veen}, S. ter and {Toribio}, C. and {Vermeulen}, R. and {Vocks}, C. and {van Weeren}, R.~J. and {Wise}, M.~W. and {Wucknitz}, O.},
        title = "{The brightness and spatial distributions of terrestrial radio sources}",
      journal = {\mnras},
         year = 2013,
        month = oct,
       volume = {435},
       number = {1},
        pages = {584-596},
          doi = {10.1093/mnras/stt1337},
archivePrefix = {arXiv},
       eprint = {1307.5580},
 primaryClass = {astro-ph.CO},
       adsurl = {https://ui.adsabs.harvard.edu/abs/2013MNRAS.435..584O}
}

@article{pankonin1981radio,
  title={Radio astronomy and spectrum management: The impact of WARC-79},
  author={Pankonin, Vwenon and Price, R Marcus},
  journal={IEEE Transactions on Electromagnetic Compatibility},
  number={3},
  pages={308--317},
  year={1981},
  publisher={IEEE},
  title={Radio astronomy and spectrum management: The impact of WARC-79},
  author={Pankonin, Vwenon and Price, R Marcus},
  journal={IEEE Transactions on Electromagnetic Compatibility},
  number={3},
  pages={308--317},
  year={1981},
  publisher={IEEE}
}

@article{galt1990contamination,
  title={Contamination from satellites},
  author={Galt, John},
  journal={Nature},
  volume={345},
  number={6275},
  pages={483--483},
  year={1990},
  publisher={Nature Publishing Group UK London}
}

@ARTICLE{5162049,
  author={Bird, Trevor S.},
  journal={IEEE Antennas and Propagation Magazine}, 
  title={Definition and Misuse of Return Loss [Report of the Transactions Editor-in-Chief]}, 
  year={2009},
  volume={51},
  number={2},
  pages={166-167},
  doi={10.1109/MAP.2009.5162049}}

@article{stone1999interference,
  title={Interference: The Limits of Radio Astronomy},
  author={Stone, W Ross},
  year={1999},
  publisher={Wiley-IEEE Press}
}

@book{thompson2017interferometry,
  author = {Thompson, A. R. and Moran, J. M. and Swenson, G. W., Jr.},
  title = {Interferometry and Synthesis in Radio Astronomy},
  edition = {3rd},
  year = {2017},
  publisher = {Springer},
address   = {Cham},
  doi = {10.1007/978-3-319-44431-4},
  note = {Includes the section "Radio Frequency Interference" (pp. 780-787)}
}

@article{pearson2016generalized,
  author = {Pearson, Ronald K. and Neuvo, Yrj{\"o} and Astola, Jaakko and Gabbouj, Moncef},
  title = {Generalized Hampel Filters},
  journal = {EURASIP Journal on Advances in Signal Processing},
  volume = {2016},
  number = {1},
  pages = {87},
  year = {2016},
  doi = {10.1186/s13634-016-0383-6},
  url = {https://doi.org/10.1186/s13634-016-0383-6}
}

@ARTICLE{2001A&A...378..327F,
       author = {{Fridman}, P.~A. and {Baan}, W.~A.},
        title = "{RFI mitigation methods in radio astronomy}",
      journal = {\aap},
         year = 2001,
        month = oct,
       volume = {378},
        pages = {327-344},
          doi = {10.1051/0004-6361:20011166},
       adsurl = {https://ui.adsabs.harvard.edu/abs/2001A&A...378..327F}
}

@ARTICLE{1965SJNA....2..205G,
       author = {{Golub}, G. and {Kahan}, W.},
        title = "{Calculating the Singular Values and Pseudo-Inverse of a Matrix}",
      journal = {SIAM Journal on Numerical Analysis},
         year = 1965,
        month = jan,
       volume = {2},
       number = {2},
        pages = {205-224},
          doi = {10.1137/0702016},
       adsurl = {https://ui.adsabs.harvard.edu/abs/1965SJNA....2..205G}
}

@ARTICLE{2018ExA....45..269S,
       author = {{Singh}, Saurabh and {Subrahmanyan}, Ravi and {Shankar}, N. Udaya and {Rao}, Mayuri Sathyanarayana and {Girish}, B.~S. and {Raghunathan}, A. and {Somashekar}, R. and {Srivani}, K.~S.},
        title = "{SARAS 2: a spectral radiometer for probing cosmic dawn and the epoch of reionization through detection of the global 21-cm signal}",
      journal = {Experimental Astronomy},
         year = 2018,
        month = apr,
       volume = {45},
       number = {2},
        pages = {269-314},
          doi = {10.1007/s10686-018-9584-3},
archivePrefix = {arXiv},
       eprint = {1710.01101},
 primaryClass = {astro-ph.IM},
       adsurl = {https://ui.adsabs.harvard.edu/abs/2018ExA....45..269S}
}

@ARTICLE{2013ExA....36..319P,
       author = {{Patra}, Nipanjana and {Subrahmanyan}, Ravi and {Raghunathan}, A. and {Udaya Shankar}, N.},
        title = "{SARAS: a precision system for measurement of the cosmic radio background and signatures from the epoch of reionization}",
      journal = {Experimental Astronomy},
         year = 2013,
        month = aug,
       volume = {36},
       number = {1-2},
        pages = {319-370},
          doi = {10.1007/s10686-013-9336-3},
archivePrefix = {arXiv},
       eprint = {1211.3800},
 primaryClass = {astro-ph.IM},
       adsurl = {https://ui.adsabs.harvard.edu/abs/2013ExA....36..319P}
}

@ARTICLE{2017ApJ...847...64M,
       author = {{Monsalve}, Raul A. and {Rogers}, Alan E.~E. and {Bowman}, Judd D. and {Mozdzen}, Thomas J.},
        title = "{Results from EDGES High-band. I. Constraints on Phenomenological Models for the Global 21 cm Signal}",
      journal = {\apj},
         year = 2017,
        month = sep,
       volume = {847},
       number = {1},
          eid = {64},
        pages = {64},
          doi = {10.3847/1538-4357/aa88d1},
archivePrefix = {arXiv},
       eprint = {1708.05817},
 primaryClass = {astro-ph.CO},
       adsurl = {https://ui.adsabs.harvard.edu/abs/2017ApJ...847...64M}
}

@ARTICLE{2022NatAs...6..984D,
       author = {{de Lera Acedo}, E. and {de Villiers}, D.~I.~L. and {Razavi-Ghods}, N. and {Handley}, W. and {Fialkov}, A. and {Magro}, A. and {Anstey}, D. and {Bevins}, H.~T.~J. and {Chiello}, R. and {Cumner}, J. and {Josaitis}, A.~T. and {Roque}, I.~L.~V. and {Sims}, P.~H. and {Scheutwinkel}, K.~H. and {Alexander}, P. and {Bernardi}, G. and {Carey}, S. and {Cavillot}, J. and {Croukamp}, W. and {Ely}, J.~A. and {Gessey-Jones}, T. and {Gueuning}, Q. and {Hills}, R. and {Kulkarni}, G. and {Maiolino}, R. and {Meerburg}, P.~D. and {Mittal}, S. and {Pritchard}, J.~R. and {Puchwein}, E. and {Saxena}, A. and {Shen}, E. and {Smirnov}, O. and {Spinelli}, M. and {Zarb-Adami}, K.},
        title = "{The REACH radiometer for detecting the 21-cm hydrogen signal from redshift z {\ensuremath{\approx}} 7.5-28}",
      journal = {Nature Astronomy},
         year = 2022,
        month = jul,
       volume = {6},
        pages = {984-998},
          doi = {10.1038/s41550-022-01709-9},
archivePrefix = {arXiv},
       eprint = {2210.07409},
 primaryClass = {astro-ph.CO},
       adsurl = {https://ui.adsabs.harvard.edu/abs/2022NatAs...6..984D}
}

@ARTICLE{2015ApJ...799...90B,
       author = {{Bernardi}, G. and {McQuinn}, M. and {Greenhill}, L.~J.},
        title = "{Foreground Model and Antenna Calibration Errors in the Measurement of the Sky-averaged {\ensuremath{\lambda}}21 cm Signal at z\raisebox{-0.5ex}\textasciitilde 20}",
      journal = {\apj},
         year = 2015,
        month = jan,
       volume = {799},
       number = {1},
          eid = {90},
        pages = {90},
          doi = {10.1088/0004-637X/799/1/90},
archivePrefix = {arXiv},
       eprint = {1404.0887},
 primaryClass = {astro-ph.CO},
       adsurl = {https://ui.adsabs.harvard.edu/abs/2015ApJ...799...90B}
}

@ARTICLE{2019JAI.....850004P,
       author = {{Philip}, L. and {Abdurashidova}, Z. and {Chiang}, H.~C. and {Ghazi}, N. and {Gumba}, A. and {Heilgendorff}, H.~M. and {J{\'a}uregui-Garc{\'\i}a}, J.~M. and {Malepe}, K. and {Nunhokee}, C.~D. and {Peterson}, J. and {Sievers}, J.~L. and {Simes}, V. and {Spann}, R.},
        title = "{Probing Radio Intensity at High-Z from Marion: 2017 Instrument}",
      journal = {Journal of Astronomical Instrumentation},
         year = 2019,
        month = jan,
       volume = {8},
       number = {2},
          eid = {1950004},
        pages = {1950004},
          doi = {10.1142/S2251171719500041},
archivePrefix = {arXiv},
       eprint = {1806.09531},
 primaryClass = {astro-ph.IM},
       adsurl = {https://ui.adsabs.harvard.edu/abs/2019JAI.....850004P}
}

@ARTICLE{2018ApJ...853..187T,
       author = {{Tauscher}, Keith and {Rapetti}, David and {Burns}, Jack O. and {Switzer}, Eric},
        title = "{Global 21 cm Signal Extraction from Foreground and Instrumental Effects. I. Pattern Recognition Framework for Separation Using Training Sets}",
      journal = {\apj},
         year = 2018,
        month = feb,
       volume = {853},
       number = {2},
          eid = {187},
        pages = {187},
          doi = {10.3847/1538-4357/aaa41f},
archivePrefix = {arXiv},
       eprint = {1711.03173},
 primaryClass = {astro-ph.IM},
       adsurl = {https://ui.adsabs.harvard.edu/abs/2018ApJ...853..187T}
}

@book{Rauscher_Spectrum_Analysis,
  author    = {Christoph Rauscher},
  title     = {Fundamentals of Spectrum Analysis},
  publisher = {{Rohde \& Schwarz GmbH \& Co. KG}},
  address   = {Munich},
  year      = {2006},
  isbn      = {3939837016}
}

@ARTICLE{Wilensky2019,
       author = {{Wilensky}, Michael J. and {Morales}, Miguel F. and
         {Hazelton}, Bryna J. and {Barry}, Nichole and {Byrne}, Ruby and
         {Roy}, Sumit},
        title = "{Absolving the SSINS of Precision Interferometric Radio Data: A New Technique for Mitigating Faint Radio Frequency Interference}",
      journal = {\pasp},
         year = 2019,
        month = nov,
       volume = {131},
       number = {1005},
        pages = {114507},
          doi = {10.1088/1538-3873/ab3cad},
archivePrefix = {arXiv},
       eprint = {1906.01093},
 primaryClass = {astro-ph.IM},
       adsurl = {https://ui.adsabs.harvard.edu/abs/2019PASP..131k4507W}
}

@ARTICLE{2023JApA...44...10B,
       author = {{Bera}, Ankita and {Ghara}, Raghunath and {Chatterjee}, Atrideb and {Datta}, Kanan K. and {Samui}, Saumyadip},
        title = "{Studying cosmic dawn using redshifted HI 21-cm signal: A brief review}",
      journal = {Journal of Astrophysics and Astronomy},
         year = 2023,
        month = jun,
       volume = {44},
       number = {1},
          eid = {10},
        pages = {10},
          doi = {10.1007/s12036-022-09904-w},
archivePrefix = {arXiv},
       eprint = {2210.12164},
 primaryClass = {astro-ph.CO},
       adsurl = {https://ui.adsabs.harvard.edu/abs/2023JApA...44...10B}
}

@ARTICLE{2023A&A...676A..75D,
       author = {{Di Vruno}, F. and {Winkel}, B. and {Bassa}, C.~G. and {J{\'o}zsa}, G.~I.~G. and {Brentjens}, M.~A. and {Jessner}, A. and {Garrington}, S.},
        title = "{Unintended electromagnetic radiation from Starlink satellites detected with LOFAR between 110 and 188 MHz}",
      journal = {\aap},
         year = 2023,
        month = aug,
       volume = {676},
          eid = {A75},
        pages = {A75},
          doi = {10.1051/0004-6361/202346374},
archivePrefix = {arXiv},
       eprint = {2307.02316},
 primaryClass = {astro-ph.IM},
       adsurl = {https://ui.adsabs.harvard.edu/abs/2023A&A...676A..75D}
}

@ARTICLE{2023ApJ...957...78W,
       author = {{Wilensky}, Michael J. and {Morales}, Miguel F. and {Hazelton}, Bryna J. and {Star}, Pyxie L. and {Barry}, Nichole and {Byrne}, Ruby and {Jordan}, C.~H. and {Jacobs}, Daniel C. and {Pober}, Jonathan C. and {Trott}, C.~M.},
        title = "{Evidence of Ultrafaint Radio Frequency Interference in Deep 21 cm Epoch of Reionization Power Spectra with the Murchison Wide-field Array}",
      journal = {\apj},
         year = 2023,
        month = nov,
       volume = {957},
       number = {2},
          eid = {78},
        pages = {78},
          doi = {10.3847/1538-4357/acffbd},
archivePrefix = {arXiv},
       eprint = {2310.03851},
 primaryClass = {astro-ph.CO},
       adsurl = {https://ui.adsabs.harvard.edu/abs/2023ApJ...957...78W}
}

@ARTICLE{2015PASA...32....8O,
       author = {{Offringa}, A.~R. and {Wayth}, R.~B. and {Hurley-Walker}, N. and {Kaplan}, D.~L. and {Barry}, N. and {Beardsley}, A.~P. and {Bell}, M.~E. and {Bernardi}, G. and {Bowman}, J.~D. and {Briggs}, F. and {Callingham}, J.~R. and {Cappallo}, R.~J. and {Carroll}, P. and {Deshpande}, A.~A. and {Dillon}, J.~S. and {Dwarakanath}, K.~S. and {Ewall-Wice}, A. and {Feng}, L. and {For}, B. -Q. and {Gaensler}, B.~M. and {Greenhill}, L.~J. and {Hancock}, P. and {Hazelton}, B.~J. and {Hewitt}, J.~N. and {Hindson}, L. and {Jacobs}, D.~C. and {Johnston-Hollitt}, M. and {Kapi{\'n}ska}, A.~D. and {Kim}, H. -S. and {Kittiwisit}, P. and {Lenc}, E. and {Line}, J. and {Loeb}, A. and {Lonsdale}, C.~J. and {McKinley}, B. and {McWhirter}, S.~R. and {Mitchell}, D.~A. and {Morales}, M.~F. and {Morgan}, E. and {Morgan}, J. and {Neben}, A.~R. and {Oberoi}, D. and {Ord}, S.~M. and {Paul}, S. and {Pindor}, B. and {Pober}, J.~C. and {Prabu}, T. and {Procopio}, P. and {Riding}, J. and {Udaya Shankar}, N. and {Sethi}, S. and {Srivani}, K.~S. and {Staveley-Smith}, L. and {Subrahmanyan}, R. and {Sullivan}, I.~S. and {Tegmark}, M. and {Thyagarajan}, N. and {Tingay}, S.~J. and {Trott}, C.~M. and {Webster}, R.~L. and {Williams}, A. and {Williams}, C.~L. and {Wu}, C. and {Wyithe}, J.~S. and {Zheng}, Q.},
        title = "{The Low-Frequency Environment of the Murchison Widefield Array: Radio-Frequency Interference Analysis and Mitigation}",
      journal = {\pasa},
         year = 2015,
        month = mar,
       volume = {32},
          eid = {e008},
        pages = {e008},
          doi = {10.1017/pasa.2015.7},
archivePrefix = {arXiv},
       eprint = {1501.03946},
 primaryClass = {astro-ph.IM},
       adsurl = {https://ui.adsabs.harvard.edu/abs/2015PASA...32....8O}
}

@ARTICLE{2023A&A...678L...6G,
       author = {{Grigg}, D. and {Tingay}, S.~J. and {Sokolowski}, M. and {Wayth}, R.~B. and {Indermuehle}, B. and {Prabu}, S.},
        title = "{Detection of intended and unintended emissions from Starlink satellites in the SKA-Low frequency range, at the SKA-Low site, with an SKA-Low station analogue}",
      journal = {\aap},
         year = 2023,
        month = oct,
       volume = {678},
          eid = {L6},
        pages = {L6},
          doi = {10.1051/0004-6361/202347654},
archivePrefix = {arXiv},
       eprint = {2309.15672},
 primaryClass = {astro-ph.IM},
       adsurl = {https://ui.adsabs.harvard.edu/abs/2023A&A...678L...6G}
}

@ARTICLE{2017AJ....153...26S,
       author = {{Sathyanarayana Rao}, Mayuri and {Subrahmanyan}, Ravi and {Udaya Shankar}, N. and {Chluba}, Jens},
        title = "{GMOSS: All-sky Model of Spectral Radio Brightness Based on Physical Components and Associated Radiative Processes}",
      journal = {\aj},
         year = 2017,
        month = jan,
       volume = {153},
       number = {1},
          eid = {26},
        pages = {26},
          doi = {10.3847/1538-3881/153/1/26},
archivePrefix = {arXiv},
       eprint = {1607.07453},
 primaryClass = {astro-ph.CO},
       adsurl = {https://ui.adsabs.harvard.edu/abs/2017AJ....153...26S}
}

\end{document}